\documentclass[twocolumn,prd,aps,amsmath,amsfonts,noshowpacs,preprintnumbers,nofootinbib]{revtex4}
\usepackage{graphicx, color}
\usepackage{subfigure}
\usepackage{tabularx}
\usepackage{longtable}
\usepackage{bbold}
\usepackage{float}
\newcolumntype{C}[1]{>{\centering\arraybackslash}b{#1}}
\newcolumntype{D}[1]{>{\centering\arraybackslash}b{#1}}
\newcolumntype{R}[1]{>{\raggedleft\arraybackslash}b{#1}}

\def\e{\mathrm{e}}
\def\p{\partial}
\def\I{\mathrm{i}}
\def\bp{\beta_+}
\def\bm{\beta_-}
\def\bpm{\beta_\pm}
\def\LB{\Delta_\mathrm{LB}}
\def\Z{\mathbb{Z}}
\def\F{\mathcal{F}}
\def\PSL{\mathrm{PSL}}
\def\PGL{\mathrm{PGL}}
\def\erf{\mathrm{erf}}
\def\nnquad{\!\!\!\!}
\def\nquad{\!\!}
\def\loOm{{\stackrel{}{\Omega}}}

\newcommand{\diffp}[2]{\frac{\p #1}{\p #2}}
\newcommand{\ddiffp}[2]{\frac{{\p}^2 #1}{{\p #2}^2}}

\newcommand{\be}{\begin{eqnarray}}
\newcommand{\ee}{\end{eqnarray}}
\newcommand{\nn}{\nonumber}
\newcommand{\bsm}{\begin{smallmatrix}}
\newcommand{\esm}{\end{smallmatrix}}

\newcommand{\dd}{{\mathrm d}}

\newcommand{\R}{\mathbb{R}}

\newcommand{\mac}[1]{\mathcal{#1}}
\newcommand{\braket}[2]{\langle #1 | #2 \rangle}

\begin{document}

\preprint{AEI-2011-050}

\title{Relativistic Wavepackets in Classically Chaotic Quantum Cosmological Billiards}

\author{Michael Koehn}
\affiliation{Max-Planck-Institut f\"ur Gravitationsphysik, Albert-Einstein-Institut, 
Am M\"uhlenberg 1, 14476 Potsdam, Germany}


\begin{abstract}

Close to a spacelike singularity, pure gravity and supergravity in four to eleven spacetime dimensions admit a cosmological billiard description based on hyperbolic Kac-Moody groups. We investigate the quantum cosmological billiards of relativistic wavepackets towards the singularity, employing flat and hyperbolic space descriptions for the quantum billiards. We find that the strongly chaotic classical billiard motion of four-dimensional pure gravity corresponds to a spreading wavepacket subject to successive redshifts and tending to zero as the singularity is approached. We discuss the possible implications of these results in the context of singularity resolution and compare them with those of known semiclassical approaches. As an aside, we obtain exact solutions for the one-dimensional relativistic quantum billiards with moving walls.                        
\\[1em]
\end{abstract}

\maketitle

\section{Introduction}
\vspace{-0.27em}

Spacetime singularities are known to generically appear in the classical theory of general relativity \cite{HP70}, and they are expected to be resolved through quantum effects. Understanding the classical structure near these singularities should shed light on the mechanisms that lead to the quantum resolution. Via the Belinskii-Khalatnikov-Lifshitz (BKL) analysis \cite{BKL70,BKL82} (see also \cite{Mis69b}), the equations of general relativity can be traced back towards the initial singularity, leading in the limit to an ultralocal description of spacetime in terms of a system of steep walls with the logarithms of the spatial scale factors being effectively described by a massless particle bouncing and specularly reflecting off these walls. In between these reflections, the universe is undergoing free Kasner evolutions \cite{RS75, Mis94}. Generally, a dynamical system of a pointlike particle moving along geodesics on a Riemannian manifold and bouncing specularly off its piecewise smooth boundary is called a billiard system, and thus the relativistic billiards arising from the BKL ana\-lysis have been termed cosmological billiards. Depending on the possible matter content and the dimension of the respective gravity theory, this motion may or may not be chaotic \cite{DHN03,HPS07}.  If it is chaotic, as is the case e.g.~for pure gravity in four and supergravity in eleven spacetime dimensions, then the motion in scale factor space undergoes an infinite series of epochs of anisotropic expansion and contraction as the singularity is approached.

While the dynamics of the gravitational field generally shows a hidden symmetry in terms of Lorentzian Kac-Moody algebras \cite{DH01}, an underlying algebraic structure of pure gravity and supergravity theories with a number of spacetime dimensions from four to eleven was also revealed through the shapes of the billiard tables which arise in the BKL limit. 
Namely, these billiard tables coincide with the fundamental Weyl chambers of infinite-dimensional Kac-Moody algebras \cite{DHN03,HPS07}, the latter being of the hyperbolic type if the corresponding billiard motion is chaotic, as is the case for all the maximally supersymmetric theories that are candidates for a unified description of the fundamental forces. Remarkably, the highest possible rank for a hyperbolic Kac-Moody algebra implies a maximum of eleven spacetime dimensions for the corresponding supergravity theory.

For theories in a certain number of spacetime dimensions, the even Weyl groups of the respective algebra have in turn been identified with generalized modular groups based on arithmetic integer rings of different kinds of division algebras \cite{FKN09,KNP10}. For example, the billiard domain corresponding to $D=4$ pure gravity is given by the fundamental Weyl chamber of the infinite-dimensional hyperbolic Kac-Moody algebra $AE_3$ \cite{DHJN01}, or equivalently by half the fundamental domain of the standard modular group on the hyperbolic upper half-plane. In the case of $D=11$ supergravity on the other hand, the relevant Kac-Moody algebra is $E_{10}$ \cite{DH01}, and the billiard table is given on the nine-dimensional generalized upper half-plane by half the fundamental domain of the generalized modular group with respect to integer octonions, see \cite{FKN11} for details on the explicit construction of such billiard domains.

For the resolution of the initial singularity of the classical gravity theory, quantum effects are expected to be the essential additional ingredient. First insights in this direction arise from the quantization of the relativistic cosmological billiards system. Following the original suggestion of \cite{Mis72}, the potential walls are translated into Dirichlet boundary conditions on the wavefunction, leading to a quantum cosmological billiard description evolving according to the free Wheeler-DeWitt equation and subjected to the boundary conditions. The generally constantly moving domain walls can be turned into static ones through a transformation to hyperbolic space, leading to an eigenvalue problem of the Laplacian on the arithmetic billiard domain (see e.g.~\cite{Iwa02} for the mathematical background in the case of complex numbers). While an exact solution of this classically chaotic quantum billiard system is unknown and out of reach, we pointed out in a previous article that the quantum cosmological billiards as a Dirichlet problem implies on the one hand the pointwise vanishing of what is commonly called the ``wavefunction of the universe'' as the singularity is approached, and furthermore that the wavefunction is generically complex and oscillating, particularly that it can not be analytically extended beyond the singularity \cite{KKN09}. In contrast to the standard Wheeler-DeWitt equation with a potential, in the quantum cosmological billiards setting it is consistent to restrict to positive norm states, but this requires the wavefunction to be complex. While complexity of the wavefunction has been proposed to be related to the emergence of a directed time in canonical gravity \cite{Bar93,Ish91,Kie07b}, it is also essential for a semiclassical wavepacket to have a definite direction of propagation.

The hyperbolic space billiard domains for $D\leq 10$ pure gravity and $D=11$ supergravity, where $D$ denotes the number of spacetime dimensions, are non-compact but of finite volume in spaces of constant negative curvature. The geodesic flow inside these domains is not only ergodic and mixing with respect to the Liouville measure, but it is furthermore uniformly hyperbolic, having the Anosov property \cite{Hop39,Ano67}. This is the strongest form of chaos, all the periodic trajectories are isolated and unstable under small variations of the initial conditions. The quantization of {\it non-relativistic} classically chaotic systems has been widely investigated, for such pure reasons as trying to understand the transition between quantum and classical mechanics for chaotic motion, and in particular to understand which imprints the nonlinear classical systems leave in their respective linear quantum counterparts, see e.g.~\cite{Gut90,Haa10}. This field of research has been termed quantum chaos, and billiard systems have served as its most popular models.

Standard semiclassical (or geometric optics) techniques are applicable only for classically completely integrable systems, and are problematic for ergodic systems due to their inconsistency with the isotropic distribution of momenta in the classical limit, as had already been noted in \cite{Ein17}. While the Einstein-Brillouin-Keller (EBK) torus quantization of integrable systems is well understood, the quantization of classically chaotic systems to date requires numerical techniques. Nevertheless, with the help of the Gutzwiller-Selberg trace formula (which reduces for integrable systems to EBK quantization), it has been shown that there exist strong correspondences of the classical system to its quantized version, relating the lengths of the classical periodic orbits to the eigenvalue spectrum \cite{Sel56,Hej7683,Gut71,Gut90}. Generally, numerically computed eigenfunctions of classically chaotic systems show regions of increased density along unstable periodic orbits, called quantum scars \cite{Hel84}. However, for the bound states inside (twice) the billiard domain for $D=4$ pure gravity, quantum unique ergodicity (QUE) has recently been proven \cite{RS94,Lin06,Sou10}. QUE implies that the eigenstates equidistribute in the semiclassical limit of high energies, excluding the possibility of localization of semiclassical eigenstates onto unstable periodic orbits, or of fractal structures.

In this article, we report investigations concerning the behavior of {\it relativistic} wavepackets evolving as quantum cosmological billiards. Along with analytical considerations (cf.~also \cite{Sch05}), we solve the corresponding Wheeler-DeWitt equation numerically, thereby obtaining detailed insights into the evolution of semiclassical wavepackets and in particular in their long-term behavior. As a prototype, we study the case of $D=4$ pure gravity, since the higher-dimensional cases are numerically too involved. We refer however to \cite{Sil05,SV07,SV11,Mar10,Mar10b} for results on QUE for higher-rank locally symmetric spaces. The main results reported in this article are as follows:
\begin{itemize}
\item The relativistic wavepackets follow their corresponding classical trajectories and then start to deviate due to transversal spreading and reflections off the moving billiard walls.\vspace{-.6em}
\item The wavefunction tends to zero as the singularity is approached.\vspace{-.6em}
\item In the flat space description of the cosmological billiards, the energy expectation value of the wavepacket decreases due to successive redshifts upon being reflected off the moving billiard walls.\vspace{-.6em}
\item In the hyperbolic space description, the energy expectation value of the wavepacket remains constant. In agreement with \cite{Mis69b}, we find that a highly excited state remains highly excited on its way into the singularity. However, the vanishing of the spreading wavepacket still suggests a quantum resolution of the singularity.\vspace{-.6em}
\item As an aside, we obtain a set of exact solutions for the relativistic quantum billiards corresponding to the one-dimensional infinite square well with moving walls, for the massless as well as for the massive case.
\end{itemize}

In section \ref{sec:wdw}, we summarize the notation and derivation of the Wheeler-DeWitt operator for the setting of quantum cosmological billiards. In section \ref{sec:1d}, we investigate the reflection of a relativistic wavepacket off a moving wall using the one-dimensional example of a moving infinite square well potential and provide a set of analytic solutions. In section \ref{sec:qcb}, we report our results concerning the quantum cosmological billiards of $D=4$ pure gravity. These have been obtained through two different approaches, namely on the one hand as a quantum billiards in flat space with moving walls and on the other hand in hyperbolic space. We then compare our findings with the predictions of \cite{KKN09}. In the appendix, we supplement these results with and contrast them to the still non-separable but integrable case of the triangular billiards in flat space with all walls kept fixed.

\section{The Wheeler-DeWitt operator for Quantum Cosmological Billiards}\label{sec:wdw}

In the BKL limit, spatial inhomogeneities and matter degrees of freedom contained in the original Lagrangian description of the gravitational theory are represented merely through the infinitely steep billiard walls \cite{DHN03,KKN09}. For the dynamics, one is left with the diagonal part of the metric, which can be written for a $D=d+1$-dimensional spacetime in pseudo-Gaussian gauge as
\be
\dd s^2 = -N(\tau)^2\dd \tau^2 + \sum_{\mu=1}^{d} \e^{-2\beta^\mu(\tau)}({\dd x^\mu})^2 \ ,
\ee
where $\mu\in\{1,\ldots, d\}$, and where $\tau$ is any global time function parametrizing the foliation into space-like hypersurfaces. We use diagonal coordinates $\beta^\mu$ in the induced space of the logarithmic scale factors, which we call $\beta$-space. In these diagonal coordinates, the Lorentzian DeWitt metric reduces to the Lorentzian minisuperspace \cite{Mis72} metric and we can write the corresponding kinetic term simply as
\be
\mathcal{L}_{\text{kin}}=\frac{1}{4n}\eta_{\mu\nu}\dot{\beta}^\mu\dot{\beta}^\nu
\ee
in terms of the rescaled lapse function $n=N/\sqrt{g}$ (see e.g.~\cite{Kie07b}) and with $\dot\beta^\mu\equiv\diffp{\beta^\mu}{\tau}$. The spatial volume collapses to zero at each spatial point at the spacelike singularity in the past ($\sqrt{g}\rightarrow 0$), where the proper time goes towards $0^+$. The scale factor $\Omega$ of the universe is related to the spatial volume $\sqrt{g}$ through $\Omega=-\frac1d\ln\sqrt{g}$. We choose the lapse for our analysis in flat $\beta$-space such that $\Omega\equiv\beta^1$ serves as the time coordinate, i.e.~the singularity is at $\Omega=+\infty$. The remaining $\beta^i$ with $i\in\{2,\ldots,d\}$ serve as the parameters of anisotropy.
Because of the spatial ultra-locality of the BKL limit, the gravitational model is reduced to the classical mechanics system of a massless relativistic billiard ball which is described by the $\beta$ variables and which moves along geodesics with respect to the flat DeWitt metric, interspersed by specular reflections off the walls. The geodesic parts of the dynamics between each two successive reflections are Kasner regimes. For each spatial point, there is one such system, and all these systems are decoupled from each other.

\begin{figure}
\includegraphics{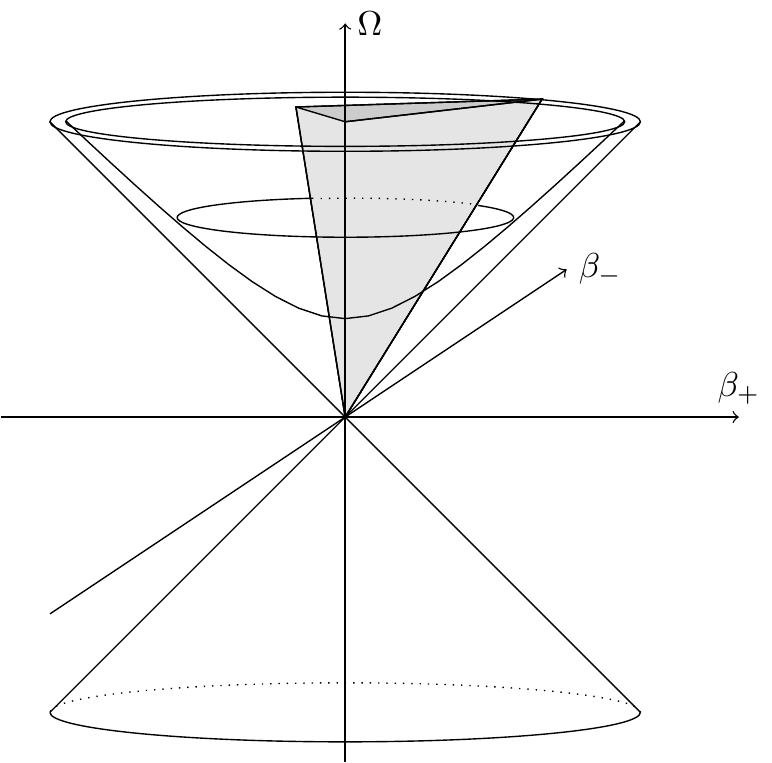}
\caption{Sketch of the cosmological billiards domain as a wedge inside the forward lightcone. It is depicted schematically how the billiard wedge is intersected by a spacelike hyperboloid.}\label{fig:kegel}
\vskip1em
\hrule\hrule
\end{figure}

With the conjugate momenta $\pi_\mu=\frac1{2}\eta_{\mu\nu}\dot\beta^\nu$, where we have set the rescaled lapse function to $n=1$, we can write the free Hamiltonian as $\mac{H}= \pi_\mu \pi^\mu$. The billiard walls are not all static with respect to $\Omega$-time in the flat Lorentzian $\beta$-space, instead some of them move with constant velocity. However, by slicing the forward lightcone of $\beta$-space in terms of hyperboloids and transforming to hyperbolic coordinates (cf.~FIG.~\ref{fig:kegel}), we obtain a billiard system with motionless walls on a negatively curved space, where we adopt the hyperbolic radial coordinate $\rho$ as the new time coordinate. Explicitly, through
\be\label{hyptraf}
\beta^\mu=\rho\gamma^\mu \quad \gamma_\mu \gamma^\mu=-1 \quad \rho^2=-\beta_\mu \beta^\mu \ ,
\ee
the Wheeler-DeWitt operator can be written in hyperbolic space as 
\be\label{WDWophyp}
\hat{\mac{H}}=\rho^{1-d}\diffp{}{\rho}\left(\rho^{d-1}\diffp{}{\rho}\right)-\rho^{-2}\LB \ .
\ee
In the hyperbolic coordinates, the singularity is at $\rho=+\infty$. We note that the factor ordering is determined through the transformation from flat space, i.e.~the whole Wheeler-DeWitt operator $\hat{\mac{H}}$ is simply the Laplace-Beltrami operator in hyperbolic coordinates, while
\be
\LB=v^{d-1}\p_v(v^{3-d}\p_v)+v^2\p_{\vec u}^2
\ee
refers to its restriction to a fixed hyperboloid. We choose the generalized upper half-plane with coordinates $\vec u\in\R^{d-2}$, $v\in\R_{>0}$ and the Poincar\'e metric $\dd\sigma^2=v^{-2}(\dd v^2+\dd{\vec u}^2)$ as a realization of the hyperbolic space. The Wheeler-DeWitt equation reads $\hat{\mac{H}}\Psi=0$ for the wavefunction $\Psi$, in flat and respectively in hyperbolic space. The infinitely steep potential walls of the billiard domain are implemented as Dirichlet boundary conditions on the wavefunction. In hyperbolic space, the Wheeler-DeWitt equation separates as
\be
\Psi(\rho,\vec u,v)=R(\rho)F(\vec u, v)
\ee
and the radial part admits the solution \cite{KKN09}
\be\label{WDWsolrad}
R_\pm(\rho)=\rho^\frac{2-d}{2}\e^{\pm\I\sqrt{k^2-(\frac{d-2}2)^2}\log\rho}
\ee
if the eigenfunctions $F$ solve the eigenvalue equation
\be
\LB F(\vec u, v) = -k^2 F(\vec u, v) \ .
\ee
For $d>2$, it is apparent from \eqref{WDWsolrad} that $\Psi\rightarrow 0$ as $\rho\rightarrow\infty$. The Dirichlet boundary conditions at the infinite potential walls that define the billiard table furthermore imply that $k^2\geq \frac{d-2}2$, which implies that the wave function is generically complex and oscillating \cite{KKN09}.

\section{Relativistic wavepackets in a one-dimensional infinite square well with a moving wall}\label{sec:1d}

The {\it non-relativistic} system of a one-dimensional infinite square well including a massive particle evolving according to the Schr\"odinger equation is one of the most elementary quantum mechanical systems. If the potential walls are however not static but moving, the situation is already much more complicated, since the system is then not separable anymore and it is difficult to find exact solutions. Results for this non-relativistic case have been obtained first by Doescher and Rice \cite{DR69}.

The {\it relativistic} moving-wall case is however a much less common object of study, and there appear interesting subtleties. Since we aim at investigating the quantum cosmological billiards scenario of a massless quantum particle evolving according to a Klein-Gordon-like Wheeler-DeWitt equation, we at first investigate the one-dimensional quantum billiards in an infinite square well with one static boundary and one boundary which is moving outward at constant velocity $\nu$. In order to match the notation of the other sections, we use $\Omega$ as the time and $\beta$ as the space coordinate. Where appropriate, we write for subsequent referring the equations in this section such that they are valid for an arbitrary number $D$ of spacetime dimensions within $4\leq D\leq 11$. We let $d=D-1$ denote the number of spatial dimensions.

The one-dimensional case considered in this section then corresponds to $d=2$. This system is described as the initial/boundary value problem (IBVP)
\be\label{KGmov}
\begin{cases}
\ddiffp{}{\Omega}\Psi (\Omega,\beta)=\ddiffp{}{\beta}\Psi(\Omega,\beta) & \text{in }\F \\
\Psi\vert_{\p\F} = 0 & \text{on }\p\F\\
\Psi(\Omega_0,\beta)=f(\beta)\quad (\p_t\Psi)(\Omega_0,\beta)=g(\beta)
\end{cases}
\ee
with the infinite square well specified by the domain $\F= [ 0 , L(\Omega) ]$ in terms of the length function $L(\Omega)=L_0+\nu(\Omega-\Omega_0)$, where $\nu$ denotes the speed of the receding wall, $0<\nu<1$, and $\Omega, \beta\in\R$. The transformation
\be\label{trafo1d}
\beta\mapsto\beta'=\frac{\beta}{L(\Omega)}
\ee
which is usually applied in such a scenario (see e.g.~\cite{MD91} for the Schr\"odinger case) is of not much help in obtaining analytic solutions for the relativistic moving-wall system because unlike the first-order Schr\"odinger case, the Klein-Gordon (KG) equation will not separate after the transformation \eqref{trafo1d}, although the walls will indeed be motionless.

However, by employing the transformation \eqref{hyptraf} to hyperbolic coordinates inside the forward lightcone in $1+1$-dimensional Minkowski space, we can obtain an infinite square well system with static boundaries.
Since the $\gamma$ coordinates are constrained to lie on the hyperboloid, one of the two coordinates is redundant. We parametrize by using coordinates on the one-dimensional analogue of the hyperbolic upper half-plane through
\be
\gamma^1=\frac{v^2+1}{2v}\qquad\gamma^2=\frac{v^2-1}{2v} \ .
\ee
In the hyperbolic coordinates, we are able to find a full set of exact solutions if we require as an intermediate step that $L_0=\nu t_0$. We may afterwards freely shift the tip of the lightcone in order to obtain solutions for general parameters $(L_0,t_0)$. After the transformation to hyperbolic space, the time dependence of the right boundary drops out and it depends merely on the constant velocity,
\be
\Lambda(\nu)=\frac{1+\nu+\sqrt{1-\nu^2}}{1-\nu+\sqrt{1-\nu^2}} \ .
\ee
The transformed massless KG equation reads
\be
\rho\p_\rho\left(\rho\p_\rho\Psi(\rho,v)\right)=v\p_v\left(v\p_v\Psi(\rho,v)\right)
\ee
and the general solution is given by
\be
\Psi(\rho,v)=\psi_1(\rho v)+\psi_2(\frac{v}{\rho}) \ ,
\ee
e.g.
\be\label{kg1dsolhyp}
\Psi_n(\rho,v)=\frac{1}{\sqrt{n\pi}}\sin\left(k_n\ln(v)\right)\exp(-\I k_n\ln(\rho))
\ee
with
\be
k_n=\frac{n\pi}{\ln(\Lambda(\nu))} \ .
\ee
We note in passing that the massive case corresponding to $\p_\Omega^2\Phi=\p_\beta^2\Phi-m^2\Phi$ can be treated similarly, leading to the solutions
\be
\Phi_n(\rho,v)=C\sin\left(k_n\ln(v)\right)\left[J_{\I k}(m\rho)+Y_{\I k}(m\rho)\right] \ ,
\ee
where $J_{\I k}$ and $Y_{\I k}$ are Bessel functions of imaginary order $\I k$.

The solutions \eqref{kg1dsolhyp} are normalized to $1$ with respect to the Klein-Gordon-like scalar product
\be\label{hypscalar}
\left(\psi_1,\psi_2\right)= \I \rho^{d-1}\int\dd \mathrm{vol}\psi_1^*\nquad\stackrel{\leftrightarrow}{\p_\rho}\nquad\psi_2 \ ,
\ee
where $\psi\nnquad\stackrel{\leftrightarrow}{\p_\rho}\nnquad\phi\equiv\psi\p_\rho\phi-\phi\p_\rho\psi$ and where $\dd\mathrm{vol}=\dd v \dd^{d-2} u v^{1-d}$ denotes the volume element on the $(d-1)$-dimensional hyperboloid.
We can express the solutions \eqref{kg1dsolhyp} in flat coordinates as
\be\label{kg1dmovsolns}
\Psi_n(\Omega,\beta)=\frac{1}{\sqrt{n\pi}}\sin\left[k_n\ln\left(\textstyle\frac{\Omega'+\beta+\sqrt{\Omega'^2-\beta^2}}{\Omega'-\beta+\sqrt{\Omega'^2-\beta^2}}\right)\right]\times\nonumber\\
\times\exp\left[-\I k_n\frac12\ln\left(\Omega'^2-\beta^2\right)\right]
\ee
with $\Omega'(\Omega)\equiv \Omega-\Omega_0+\frac{L_0}{\nu}$, $0<\nu<1$. The solutions \eqref{kg1dmovsolns} are normalized to $1$ with respect to the standard form of the Klein-Gordon-invariant scalar product,
\be\label{KGscalar}
\braket{\psi_1}{\psi_2}=\I\int \dd^{d-1}\beta\psi_1^* \nquad\stackrel{\leftrightarrow}{\p_\loOm}\nquad \psi_2 \ .
\ee
Furthermore, the respective norms are preserved in hyperbolic as well as in flat space. The set of solutions \eqref{kg1dsolhyp} and \eqref{kg1dmovsolns} seems to be original to us, see however the appendix of \cite{Rya72} for first attempts in the treatment of the relativistic one-dimensional moving-wall system.

\begin{figure*}[!ht]
\subfigure[Before reflection]{\includegraphics[width=0.24\textwidth]{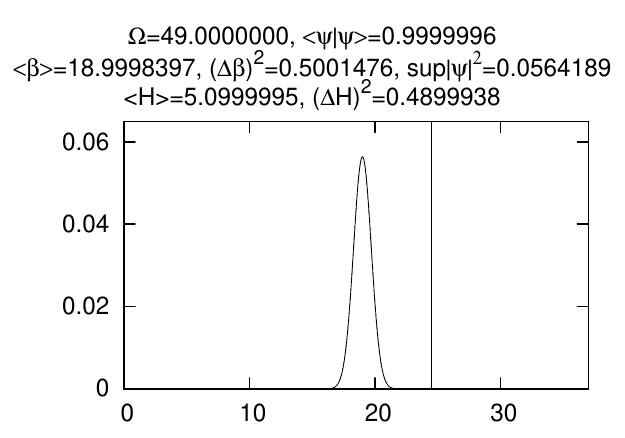}}
\subfigure[During reflection]{\includegraphics[width=0.24\textwidth]{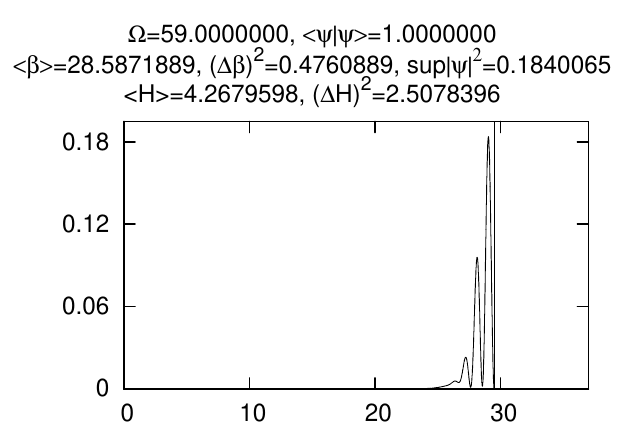}}
\subfigure[Redshifted wavepacket]{\includegraphics[width=0.24\textwidth]{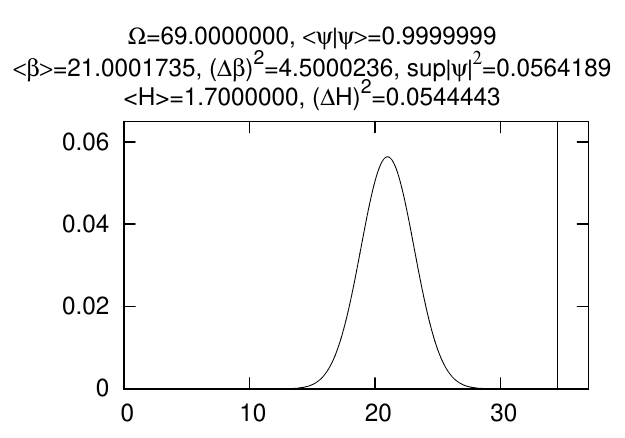}}
\subfigure[Late-time behavior (scale changed)]{\includegraphics[width=0.24\textwidth]{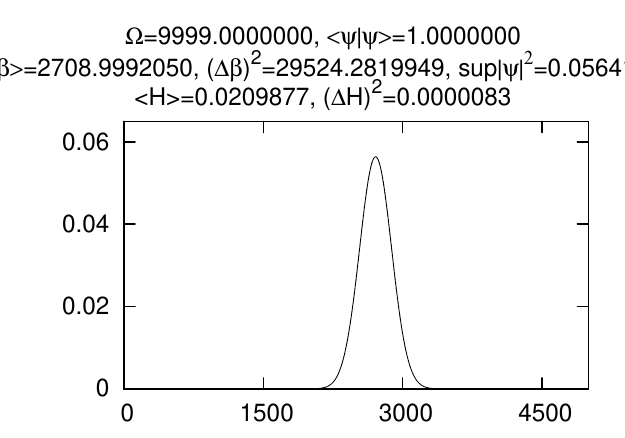}}
\caption{Plots of $|\psi(\beta)|^2$ of a one-dimensional Gaussian wavepacket redshifted upon reflection off a moving wall in flat space. The vertical bar in (a)-(c) represents the moving wall. The horizontal scaling in (d) differs from (a)-(c).}\label{fig:1dmov}
\end{figure*}

\begin{figure*}[!ht]
\subfigure[Before reflection]{\includegraphics[width=0.24\textwidth]{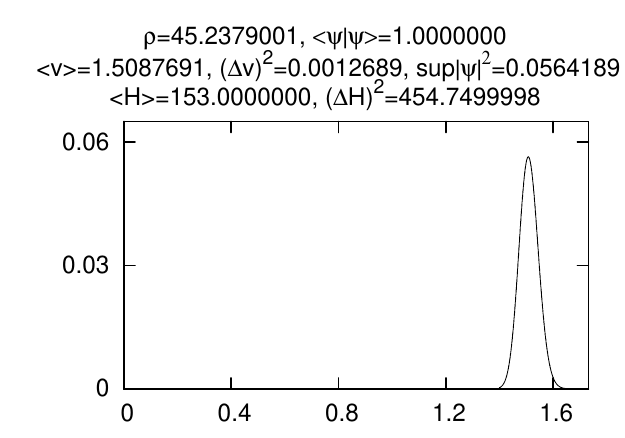}}
\subfigure[During reflection]{\includegraphics[width=0.24\textwidth]{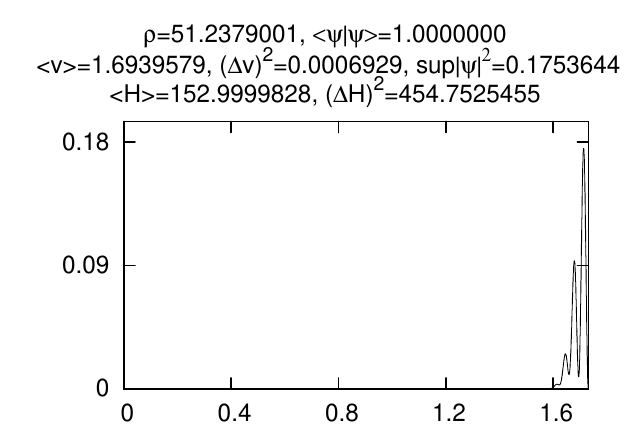}}
\subfigure[Reflected wavepacket]{\includegraphics[width=0.24\textwidth]{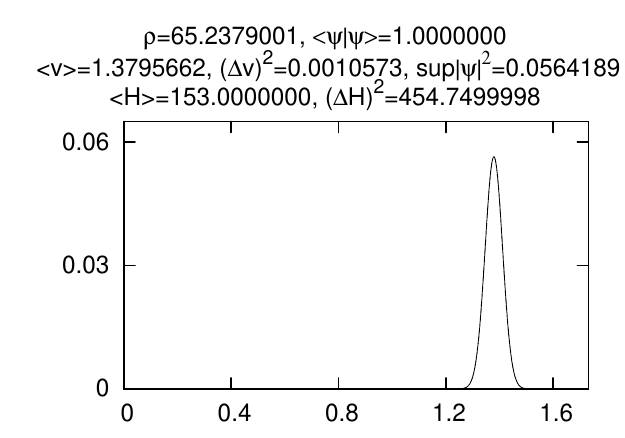}}
\subfigure[Late-time behavior]{\includegraphics[width=0.24\textwidth]{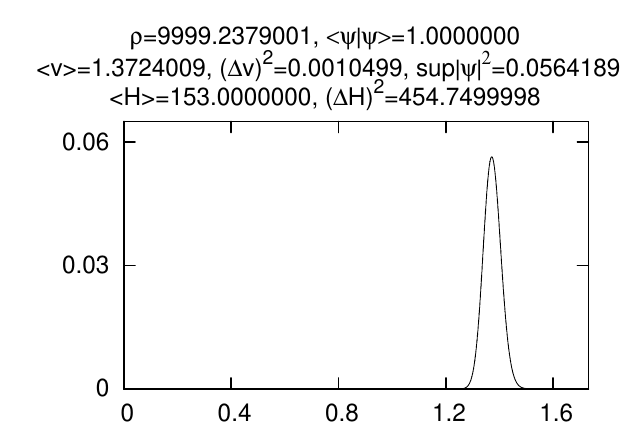}}
\caption{Plots of $|\psi(v)|^2$ of a one-dimensional Gaussian wavepacket in an infinite square well in hyperbolic space.}\label{fig:1dhyp}
\vskip1em
\hrule\hrule
\end{figure*}

For the set of solutions \eqref{kg1dmovsolns} we can directly show that
\be
\braket{\Psi_n}{\Psi_m}=\delta_{nm} \ .
\ee
A one-dimensional relativistic Gaussian wavepacket
\be\label{1pac}
\Psi(\Omega,\beta)=A\int_{-\infty}^\infty\dd p \e^{-\frac{c^2(p-p_0)^2}2+\I(p\beta-\omega \Omega)}
\ee
composed out of positive frequencies $\omega=|p|$ will have a norm of one with respect to \eqref{KGscalar} if
\be
A=\frac{c}{\sqrt\pi}\left( \e^{-c^2p_0^2}+\sqrt\pi p_0c\ \erf(p_0c) \right)^{-\frac12}\ ,
\ee
where $\erf$ denotes the error function. The wavepacket \eqref{1pac} has constant norm (of approximately $1$ if $c$ is chosen small enough) with respect to \eqref{KGscalar} during its evolution in the moving wall system, however its absolute width with respect to the $\beta$-coordinate will grow with each reflection off the moving wall, as the wavepacket will go through successive {\it redshifts}. If a one-dimensional wave reflects off a wall which is moving away at a constant speed, it experiences a redshift given by
\be\label{redshift}
1+z=\frac{f_\mathrm{ref}}{f_\mathrm{inc}}=\gamma^2 (1+\nu)^2=\frac{1+\nu}{1-\nu} \ ,
\ee
with $\gamma=(1-\nu^2)^{-\frac12}$, e.g.~for $\nu=\frac12$, the redshift is $1+z=3$. In terms of the reduced Hamiltonian $H=\sqrt{{\pi_\beta}^2}$, where $2\mac{H}=-\pi_\Omega^2+H^2$, the corresponding reflection of a classical massless particle in one dimension off a moving wall implies the relation
\be
1+z=\frac{H_\mathrm{ref}}{H_\mathrm{inc}} \ ,
\ee
which analogously simply states the relation of the energy $H_\mathrm{inc}$ of a photon before a bounce from a moving mirror to its energy $H_\mathrm{ref}$ afterwards.
As a measure for the redshift, we employ the expectation value $\langle \hat H \rangle$ of the reduced Hamiltonian operator, which we call the energy expectation value in the following. In order to avoid the square root in the reduced Hamiltonian, especially in view of the higher-dimensional cases, we adopt the two-component notation \cite{FV58} for \eqref{KGmov} by defining
\be
\psi=\phi+\chi\quad \I\p_\loOm\psi=\phi-\chi
\ee
and adapt it to the massless case
to obtain
\be\label{Hsqrt}
\hat H=-\frac{\sigma_3+\I\sigma_2}{2}\Delta+\frac{\sigma_3-\I\sigma_2}{2} \ ,
\ee
where $\sigma_i$ are the Pauli matrices and where $\Delta$ is the Laplace operator in flat space. Upon defining
\be
\Psi=\left(\bsm \phi \\ \chi \esm\right)\ ,
\ee
equation \eqref{KGscalar} is expressed through
\be\label{KGvector}
\langle \Psi_1 | \Psi_2 \rangle = 2 \int \dd^{d-1}\beta \Psi_1^\dagger \sigma_3 \Psi_2
\ee
and the energy expectation value through
\be\label{energy1d}
\langle \hat{H} \rangle = 2\int \dd^{d-1}\beta \Psi^\dagger \sigma_3 \hat{H}\Psi \ .
\ee
The two-component form \eqref{Hsqrt} implies the expected expression
\be
\langle \hat{H}^2 \rangle = -\I\int \dd^{d-1}\beta \psi^* \nquad\stackrel{\leftrightarrow}{\p_\loOm}\nquad (\Delta\psi)
\ee
for the expectation value of the squared reduced Hamiltonian. Upon reflection off a moving wall in flat space, the wavepacket loses energy into the wall, while its KG norm is preserved. See FIG.~\ref{fig:1dmov} for an illustration of such a bounce, calculated by numerically solving the KG equation on the one hand in flat space with moving boundary conditions, and on the other hand an illustration of the corresponding evolution in hyperbolic space with static boundary conditions in FIG.~\ref{fig:1dhyp}. In order to observe the level excitation distribution for the bound problem in hyperbolic space, we set
\be
\langle \hat{H_\rho}^2 \rangle = -\I\rho^{d-1}\int \dd \mathrm{vol} \psi^* \nquad\stackrel{\leftrightarrow}{\p_\rho}\nquad (\Delta_\mathrm{LB}\psi) \ ,
\ee
and thus the expectation value of the reduced Hamiltonian on the hyperbolic plane in two-component notation as
\be\label{hypenergy}
\langle \hat{H}_\rho \rangle = 2\int \dd\mathrm{vol} \Psi^\dagger \sigma_3 \hat{H}_\rho\Psi
\ee
with
\be
\hat{H}_\rho=-\rho^{d-2}\frac{\sigma_3+\I\sigma_2}{2}\Delta_\mathrm{LB}+\rho^{2-d}\frac{\sigma_3-\I\sigma_2}{2} \ .
\ee
For the derivation we have set
\be
\psi=\phi+\chi\quad \I\rho^{d-1}\p_\rho\psi=\phi-\chi \ .
\ee

Note furthermore that in the one-dimensional case considered in this section, the supremum norm $|\psi|_\infty$ of the free wavepacket remains constant, in accordance with the solution \eqref{WDWsolrad} to the radial part of the $d$-dimensional Wheeler-DeWitt equation for $d=2$.

Using \eqref{KGvector}, we compute position expectation values according to
\be\label{posex}
\langle \hat \beta_i \rangle = 2 \int\dd^{d-1}\beta\Psi^\dagger \sigma_3 \beta_i \Psi = \I \int \dd^{d-1}\beta \beta_i\psi^* \nquad\stackrel{\leftrightarrow}{\p_\loOm}\nquad \psi
\ee
and analogously in hyperbolic space using \eqref{hypscalar}. We remark however that, unlike the Hamiltonian and the momentum operator, the position operator in relativistic quantum mechanics generally mixes positive and negative frequency components of the wavepacket \cite{FV58}. We nevertheless adopt the prescription \eqref{posex} as the intuitive measure for the ``center of mass'' of the wavepacket and employ the usual Klein-Gordon covariant wavepackets (cf. e.g.~\eqref{gensol} below). For details on localization of relativistic particles, we refer the reader to the investigations in \cite{NW49,Wig62}, and to \cite{HO93} in the setting of quantum cosmology.

\section{Quantum Cosmological Billiards}\label{sec:qcb}

The classically strongly chaotic quantum BKL billiards corresponding to $D=4$ pure gravity serves as the prototype for studying the relativistic wave chaos inherent also in the relevant higher-dimensional theories (see e.g.~\cite{DHN03}). If $D>4$, the situation becomes quickly unfeasible for numerical studies. In the case of $D=4$ pure gravity, the billiard table is given by a wedge in three-dimensional flat Lorentzian $\beta$-space, cf.~FIG.~\ref{fig:kegel}, or equivalently by a triangle with one of its three edges moving outwards at constant speed $\nu$ as the singularity is approached, see FIG.~\ref{fig:cbflat}. It is not known how to solve such a triangular two-dimensional moving wall system analytically. Although by a suitable transformation the moving triangle walls in two-dimensional flat space can be transformed to static ones in hyperbolic space, the resulting static domain will be of too complicated shape. Thus we resort to numerical methods for our investigations. Since for the numerical treatment we do not have to rely on separability of the Wheeler-DeWitt equation, we studied the quantum cosmological billiards in flat as well as in hyperbolic space and confirmed that the results match.

\subsection{Flat space description}

\begin{figure}[t]
\subfigure[The cosmological billiard domain for $D=4$ pure gravity in flat space, here $\Omega_1=32$ and $\Omega_0=24$. The displayed trajectory corresponds to a classical particle with initial conditions according to \eqref{initparams}. The particle starts at the small depicted circle and is at first reflected perpendicularly from the receding wall at $a$ and then passing again through the starting point. Specular reflection does generally only hold in the rest frame of the respective wall. After the last depicted bounce at $c$, the classical particle is moving exactly parallel to the $\bp$ axis and will never hit a wall again.]{
\includegraphics{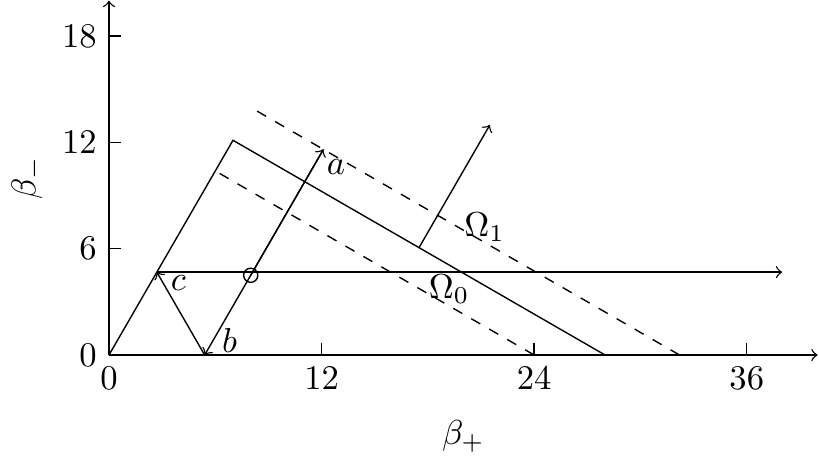}\label{fig:cbflat}}
\subfigure[The wavepacket's position expectation values according to \eqref{posex} are centered on the classical trajectory (cf.~FIG.~\ref{fig:cbflat}) in the beginning and then deviate as the wavepacket spreads and reflects. This is an enlarged excerpt of FIG.~\ref{fig:trv05-eparam}.]{\includegraphics[width=\columnwidth]{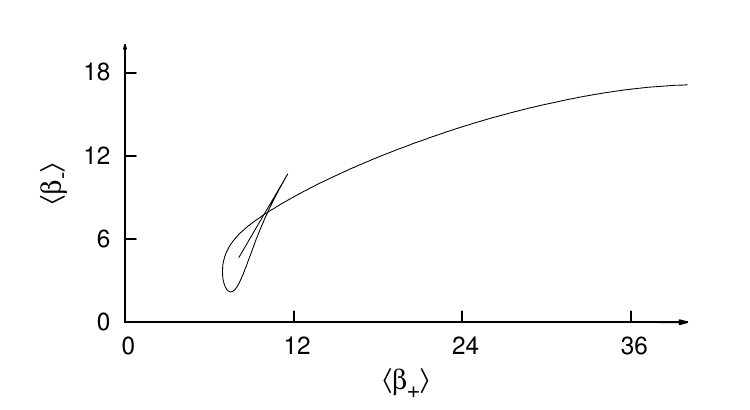}\label{fig:trv05-ep-small}}
\caption{Classical and wavepacket trajectories in the flat space description of the quantum cosmological billiards according to initial parameters \eqref{initparams}.}
\vskip1em
\hrule\hrule
\end{figure}

For the case of $d=3$, we choose the convenient standard notation $\Omega\equiv\beta^1$, $\bp\equiv\beta^2$, $\bm\equiv\beta^3$ for the coordinates. The Wheeler-DeWitt operator corresponding to \eqref{WDWophyp} then takes the form
\be
\hat{\mac{H}}\equiv \eta^{\mu\nu}\p_\mu\p_\nu=\ddiffp{}{\Omega}-\ddiffp{}{\bp}-\ddiffp{}{\bm} \ ,
\ee
and the IBVP of the quantum cosmological billiards is defined by
\be\label{KGqcbflat}
\begin{cases}
\hat{\mac{H}}\Psi(\Omega,\bpm)=0 & \text{in }\F \\
\Psi\vert_{\p\F} = 0 & \text{on }\p\F\\
\Psi(\Omega_0,\bpm)=f(\bpm) & \\
(\p_\Omega\Psi)(\Omega_0,\bpm)=g(\bpm) & 
\end{cases}
\ee
with $\F= \{ \vec\beta\in\R^2 | \bm > 0,\bm < \sqrt{3}\bp,\bm < \frac{1}{\sqrt 3}(-\bp+\Omega) \}$ representing the triangular domain with one face of the triangle moving outwards at a constant speed $\nu=\left[\sum_{i=1}^2(\diffp{\beta^i}{\Omega})^2\right]^{\frac12}=\frac12$, as displayed in FIG.~\ref{fig:cbflat}. The derivation of the wall forms is detailed in \cite{DHN03,HPS07}. Note that this triangular domain is actually the sixth part of the Mixmaster domain of a Bianchi IX universe \cite{Mis69}.

We rotate the cosmological billiard triangle counterclockwise by $\alpha=\frac{\pi}{6}$ and perform the time-dependent rescalings
\be\label{qcbrescale}
\bp'=\frac{\bp}{L_+(\Omega)}  \qquad \bm'=\frac{\bm}{L_-(\Omega)}
\ee
using the length functions $L_+(\Omega)=\frac{\sqrt3}2\Omega$ and $L_-(\Omega)=\frac12\Omega$. In the rescaled coordinates $(\bp',\bm')$, the quantum billiards is defined in the static domain given by the upper-left half of a square with side length $1$, with the Wheeler-DeWitt operator modified accordingly. The upper-right corner is constrained to be null on the forward lightcone, i.e.~it represents the cusp of the triangle on the hyperbolic upper half-plane, cf.~the following subsection.

The Wheeler-DeWitt equation is a second order hyperbolic PDE, thus we require two initial conditions. We take a plain Gaussian covariant KG wavepacket and its time derivative, chosen such as to restrict to only positive frequencies. Since the treatment will in any case be limited by the numerical accuracy, the wavepacket will fit to any desired accuracy into the box if its width $c$ is chosen small enough. A free two-dimensional Gaussian wavepacket composed out of positive-frequency modes is given by
\be\label{gensol}
\Psi(\Omega,\bpm)=A\int\limits_{-\infty}^{\infty}\dd^2p \e^{-\frac{c^2}{2}(\vec p - {\vec p}_0)^2+\I \vec p\cdot (\vec \beta -\vec\beta_0)-\I \omega (\Omega-\Omega_0)}
\ee
with ${\omega (\vec p)=\sqrt{{p_+}^2+{p_-}^2}}$ expressing the relativistic dispersion relation. Using \eqref{qcbrescale}, we can write the two initial conditions at $\Omega_0$ as
\be
\Psi\vert_{\Omega_0} & = & A\int_{-\infty}^{\infty}\dd^2p \e^{-\frac{c^2}{2}(\vec p - {\vec p}_0)^2+\I \vec p\cdot (\vec \beta -\vec\beta_0)} \label{iv1st}\\
(\p_\Omega\Psi)\vert_{\Omega_0} & = & -\I A\int_{-\infty}^{\infty}\dd^2p [\omega(\vec p) -p_+\dot\bp-p_-\dot\bm ] \times \nn\\
& & \qquad\qquad\times\ \e^{-\frac{c^2}{2}(\vec p - {\vec p}_0)^2+\I \vec p\cdot (\vec \beta -\vec\beta_0)} \ , \label{iv2nd}
\ee
where here $\dot\beta\equiv\diffp{\beta}{\Omega}$ and where the minus sign in \eqref{iv2nd} arises from restricting to positive-frequency modes. The condition \eqref{iv1st} can easily be integrated analytically, resulting in
\be
\Psi\vert_{\Omega_0}=\frac{2\pi}{c^2}A\e^{-\frac{\bp^2+\bm^2}{2 c^2}+\I \vec{p}_0(\vec\beta-\vec\beta_0)} \ ,
\ee
however \eqref{iv2nd} is not separable in the momentum variables and for $\vec p_0 \neq 0$ has to be evaluated numerically, e.g.~through Simpson's rule. Requiring the wave packet to have unit norm with respect to the KG inner product \eqref{KGscalar}
fixes the normalization constant to
\be
A & = & \frac12c^\frac12\pi^{-\frac74}\e^{\frac{c^2}{4}{\vec{p}_0}^{\,2}} \times \nn\\
& & \times \left[ \left({\textstyle c^{-2}+{\vec{p}_0}^{\,2}}\right)I_0({\textstyle\frac{c^2}{2}{\vec{p}_0}^{\,2}})+{\vec{p}_0}^{\,2}I_1({\textstyle\frac{c^2}{2}{\vec{p}_0}^{\,2}}) \right]^{-\frac12}\label{normconst}  ,
\ee
where $I_k$ is the modified Bessel function of the first kind and of order $k$.

The direction of $\vec p_0$ determines the direction of propagation of the wavepacket. Unlike the non-relativistic Schr\"odinger case, the magnitude of $\vec p_0$ does naturally not determine the wavepacket's speed of propagation, which is fixed to $1$, however it does determine its modulation frequency, i.e.~its energy, and it effects the rate of its transversal spreading. Since we are considering a relativistic wavepacket, we have to deal with a dispersion relation that states a linear relation between the absolute momentum and the frequency, unlike the quadratic relation in the Schr\"odinger case. This implies that the group speed equals the phase speed and that there is no dispersion of the wavepacket in the direction of propagation, however there will be a spreading transversally to the direction of propagation.
Setting higher values for $\vec p_0$ while keeping $c$ fixed means generating a packet which will spread more slowly.

If a plane wave hits perpendicularly a wall which is moving away at a speed $\nu_\mathrm{wall}$, it will experience a redshift $1+z$ as in \eqref{redshift}.
As in the one-dimensional case, if $\nu_\mathrm{wall}=\frac{1}{2}$, then $1+z=3$. This also holds for the classical description of the bounce, $\frac{H_\mathrm{inc}}{H_\mathrm{ref}}=3$. However, the quantum wavepacket is composed out of a superposition of plane waves with slightly non-parallel wave vectors, representing its transversal localization and fanning. Thus, in the wavepacket analogy of a classical perpendicular bounce, almost all plane wave components are in fact not hitting the wall orthogonally, implying that the energy will actually be reduced by a factor of less than three. For example, for an orthogonal bounce of a wavepacket with the initial conditions \eqref{initparams}, this results in
$\frac{\langle \hat{H}\rangle_\mathrm{inc}}{\langle \hat{H}\rangle_\mathrm{ref}}\approx 2.89$. In the limit of large momenta, we then recover the classical result of three.

Specular reflection does not hold for the massless particle bouncing off a moving wall. For non-orthogonal such bounces, we recall the relations for a classical pointlike particle \cite{Mis69b}
\be
\frac{H_\mathrm{inc}}{H_\mathrm{ref}} & = & \frac{\sin(\theta_\mathrm{ref})}{\sin(\theta_\mathrm{inc})}\\
\sin(\theta_\mathrm{ref})-\sin(\theta_\mathrm{inc}) & = & \frac12\sin(\theta_\mathrm{ref}+\theta_\mathrm{inc}) \ ,
\ee
where $\theta_\mathrm{inc}$ and $\theta_\mathrm{ref}$ denote the angles of incidence and reflection, respectively. For example, if $\theta_\mathrm{inc}=\frac{\pi}{6}$, then $\frac{H_\mathrm{inc}}{H_\mathrm{ref}}\approx 1.953$, while for a wavepacket with $|\vec p_0|=5$, we obtain $\frac{\langle \hat{H}\rangle_\mathrm{inc}}{\langle \hat{H}\rangle_\mathrm{ref}}\approx 1.97$ if it approaches the moving wall at an angle of incidence $\theta_\mathrm{inc}$.

For the numerical investigation of the quantum cosmological billiards, we need to specify the parameters for the initial data \eqref{iv1st} which are to be evolved by \eqref{KGqcbflat}. The size of the initial billiard domain and the modulation frequency of the wavepacket effect the necessary grid size of the numerical computation. Computational efficiency therefore suggests to choose the parameters for the initial conditions such that the initial billiard domain is just large enough for the initial wavepacket \eqref{iv1st} with width $c=1$ to fit in up to numerical accuracy. One possible choice, which we adopt for our simulation, is
\be\label{initparams}
\Omega_0=24 \ ,\quad {\bp}_0=8 \ ,\quad {\bm}_0=\textstyle\frac92 \ ,\quad \vec p_0= R\left(\!\begin{smallmatrix} 0 \\ 5\end{smallmatrix}\!\right) \ ,
\ee
where $R=\frac12\left(\!\begin{smallmatrix}\sqrt3 & 1 \\ -1 & \sqrt{3}\!\end{smallmatrix}\right)$ rotates clockwise by $\frac\pi 6$. The ensuing wavepacket evolution corresponds to the classical path depicted in FIG.~\ref{fig:cbflat}. We have set for our investigations $|\vec p_0|= 5$ because the momentum is on the one hand large enough such that the Gaussian momentum distribution is only negligibly indefinite, and on the other hand a low initial momentum makes numerical calculations easier because the necessary grid resolution scales with the modulation frequency squared. In the end, we can expect the results of our investigations to hold qualitatively also for larger and differently directed initial momentum choices as well as larger initial billiard domains.
We studied the evolution of this two-dimensional relativistic wavepacket numerically by specifying a two-dimensional grid. The reported results have been obtained using a grid size of $5536\times 3200$. A three-level convergence test including the additional grid sizes $4152\times 2400$ and $2768\times 1600$ ensured us about the convergence of the numerical solution (see e.g.~\cite{Ric10,Cho91}).
The Courant-Friedrichs-Lewy (CFL) condition (see e.g.~\cite{CFL28,GKO95}) states that the numerical domain of dependence must include the analytical one, implying $\Delta \Omega \leq \sigma \Delta\beta$ with the Courant number $\sigma\leq 1$. We have set $\sigma=\frac16$, and thus $\Delta\Omega=0.0025$, and increased $\Delta\Omega$ linearly according to the change of size of the billiard domain. We used fourth order classical Runge-Kutta time integration in conjunction with the method of lines (see e.g.~\cite{Sch91}) and implemented the spatial derivative operators as fourth order finite difference approximations (see e.g.~\cite{AS72}).

\begin{figure*}
\subfigure[Initial circular Gaussian wavepacket]{\includegraphics[width=\columnwidth]{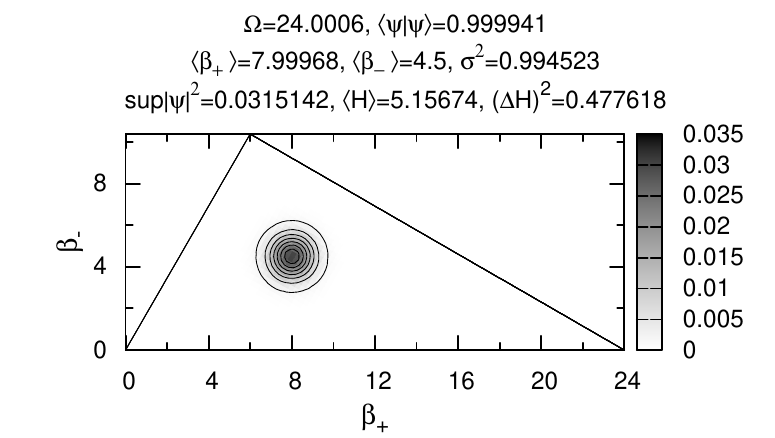}}
\subfigure[The wavepacket near point $a$ (cf.~FIG.~\ref{fig:cbflat}).]{\includegraphics[width=\columnwidth]{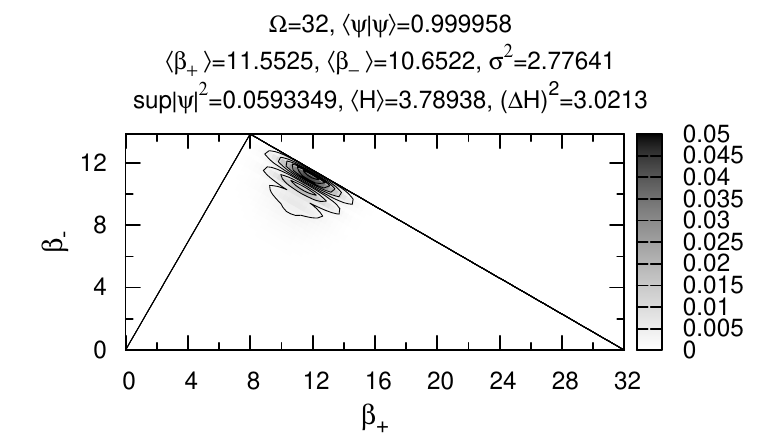}}
\subfigure[The wavepacket near point $c$ (cf.~FIG.~\ref{fig:cbflat}).]{\includegraphics[width=\columnwidth]{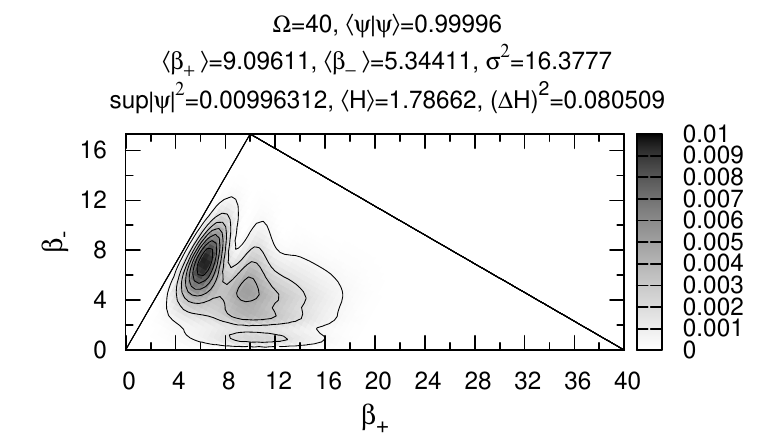}}
\subfigure[The wavepacket is moving towards the cusp while spreading.]{\includegraphics[width=\columnwidth]{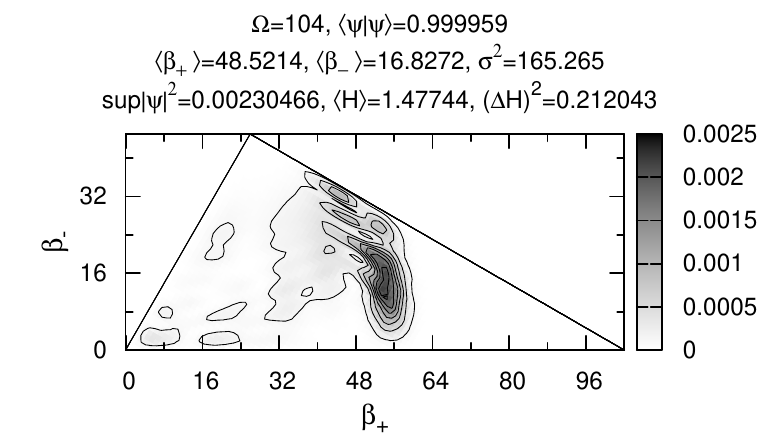}}
\subfigure[At large $\Omega$, the wavepacket has spread over the billiard domain.]{\includegraphics[width=\columnwidth]{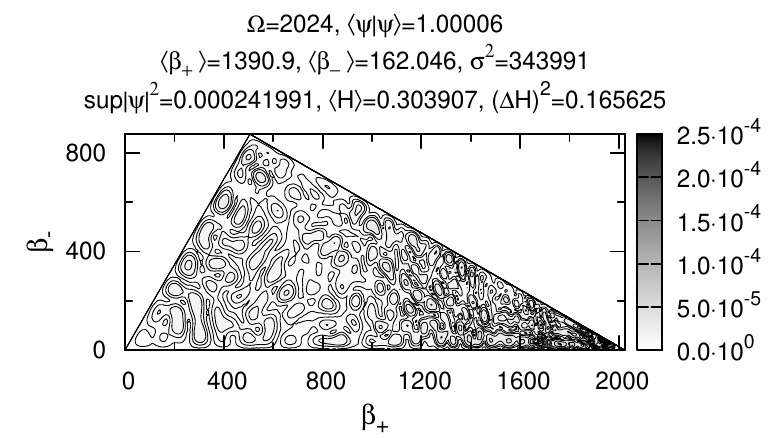}\label{fig:trv05-1000}}
\subfigure[Magnification of the cusp in FIG.~\ref{fig:trv05-1000}. Since the right corner is on the lightcone, the wavepacket can never reach it.]{\includegraphics[width=\columnwidth]{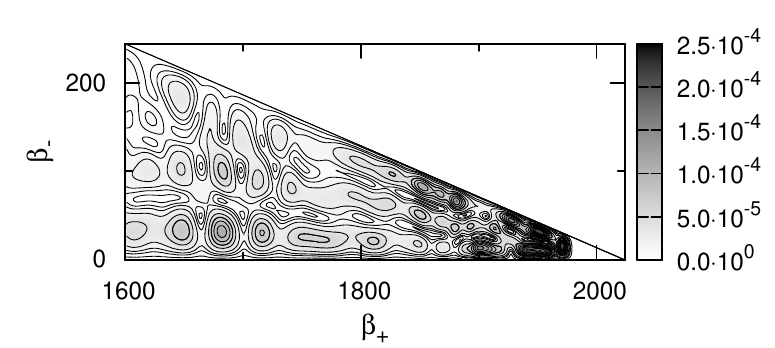}}
\caption{Snapshots of wavepacket evolution in the quantum cosmological billiards of $D=4$ pure gravity in flat space.}\label{fig:trv05-snap}
\vskip1em
\hrule\hrule
\end{figure*}

\begin{figure*}
\subfigure[The norm of the wavepacket during its evolution according to \eqref{KGscalar}. The parabolic envelope is due to the enlarging grid spacing, the oscillations mainly stem from interpolation artefacts in the computation of the $\Omega$ derivative.]{\includegraphics[width=\columnwidth]{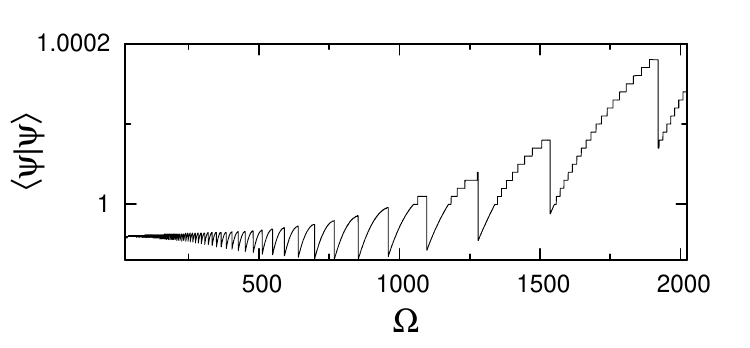}\label{fig:trv05-norm}}
\subfigure[Absolute supremum of $|\psi|^2$]{\includegraphics[width=\columnwidth]{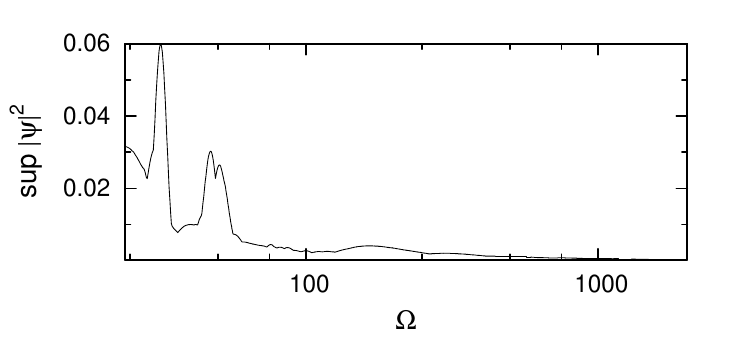}\label{fig:trv05-sup}}
\subfigure[Extended parametric plot of the of the position expectation values according to \eqref{posex}, cf.~FIG.~\ref{fig:trv05-ep-small}.]{\includegraphics[width=\columnwidth]{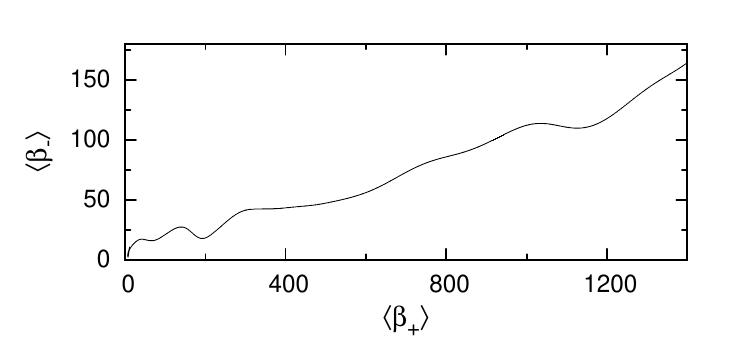}\label{fig:trv05-eparam}}
\subfigure[Long-term behavior of the expectation values of the anisotropy parameters]{\includegraphics[width=\columnwidth]{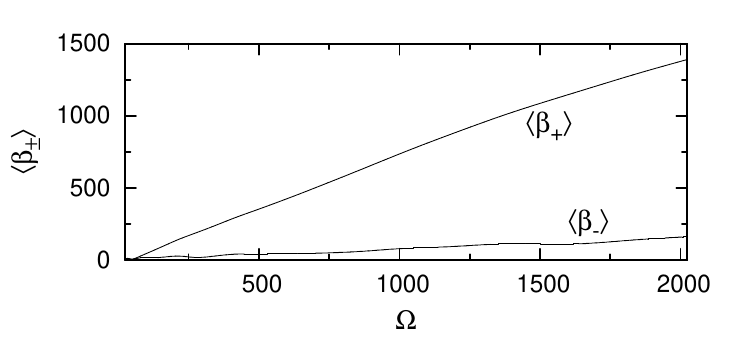}\label{fig:trv05-expect}}
\subfigure[Variance of the position expectation value]{\includegraphics[width=\columnwidth]{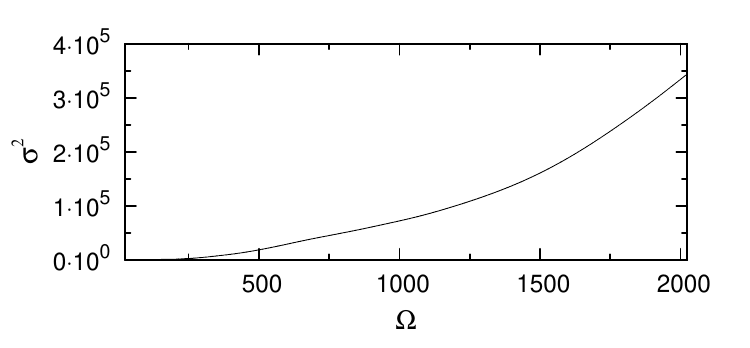}\label{fig:trv05-var}}
\subfigure[Energy expectation value of the wavepacket]{\includegraphics[width=\columnwidth]{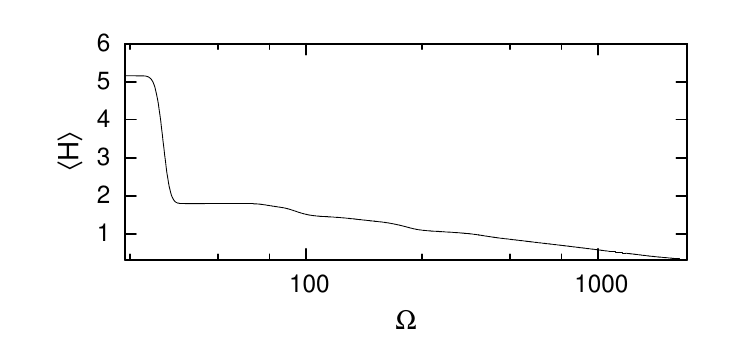}\label{fig:trv05-en}}
\subfigure[Variance of the energy expectation value]{\includegraphics[width=\columnwidth]{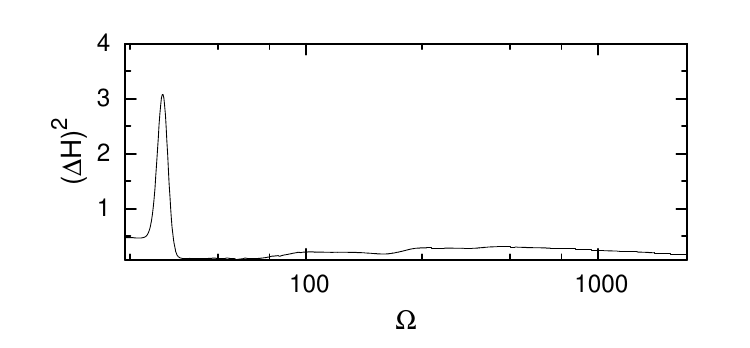}\label{fig:trv05-envar}}
\subfigure[The product of energy expectation value and time]{\includegraphics[width=\columnwidth]{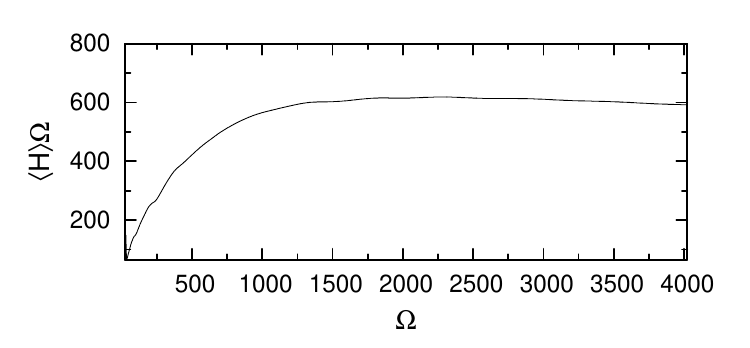}\label{fig:trv05-ent}}
\caption{Data of the wavepacket evolution with initial values \eqref{initparams} in the quantum billiards of $D=4$ pure gravity in flat space.
}\label{fig:trv05-data}
\vskip1em
\hrule\hrule
\end{figure*}

Snapshots of the evolution are displayed in FIG.~\ref{fig:trv05-snap}. As plotted in FIG.~\ref{fig:trv05-sup}, the supremum norm tends to zero with growing $\Omega$, matching the predictions of \cite{KKN09}. Indeed, with increasing $\Omega$ the wavepacket has increasingly more space to spread about. We contrast this to the case of a static triangular quantum billiards in the appendix. We furthermore computed norm, variance and expectation values according to \eqref{KGscalar} and \eqref{posex}, where we relied on Lagrange interpolation of the spatial grid for the computation of the $\Omega$ derivative. The plot of the variance in FIG.~\ref{fig:trv05-var} suggests growth of the variance without bounds as the billiard domain keeps increasing, i.e.~the wavepacket is spreading out more and more, on the one hand because of the transversal fanning, on the other hand because of the reflections off the moving wall.

The Klein-Gordon norm \eqref{KGscalar} stays constant within bounds of $3\cdot 10^{-4}$, cf.~FIG.~\ref{fig:trv05-norm}. For the chosen initial parameters, the position (i.e.~anisotropy) expectation values computed according to \eqref{posex} increase towards the cusp as the singularity is approached, as plotted in FIG.~\ref{fig:trv05-expect}. However, this can be seen to happen because the corner of the triangle which corresponds to the cusp on the hyperbolic upper half-plane is on the lightcone and thus the part of the wavepacket moving into this direction will only slowly start to reflect from the moving wall and also spread more slowly, thereby gaining in relative weight in the anisotropy expectation value computation. We furthermore observe by comparison of FIG.~\ref{fig:trv05-ep-small} with FIG.~\ref{fig:cbflat} that the wavepacket's trajectory in terms of its anisotropy expectation values \eqref{posex} follows at first the classical path and then deviates due to the spreading of the wavepacket towards the walls.

The energy expectation value and variance are given through \eqref{energy1d}. From the data displayed in FIG.~\ref{fig:trv05-en}, one observes the decrease in energy due to the wavepacket getting redshifted successively. In the following, we compare our results to the semiclassical considerations of \cite{Mis69b} for the Mixmaster universe. Although the billiard domain considered here is only the sixth part of the Mixmaster billiard domain, we can nevertheless relate the respective results through the observation that the Mixmaster domain can be folded three times along its three static heights to match the billiard domain considered here. Though a comparison should reveal a difference in the rate of decrease of the supremum norm and in the rate of increase of the position expectation values and variance, we can expect the results about the rate of energy loss to be the same in both cases. Through an adiabatic approximation, Misner examined in \cite{Mis69b} on the one hand the bound states of the billiard triangle for a given instant of time $\Omega$ and on the other hand the classical trajectories of massless particles and the energy loss involved from reflections off moving walls. From this analysis it turns out that the product of energy expectation value and $\Omega$ is roughly invariant, and thus that a semiclassical, highly excited state has to remain highly excited all the way into the singularity. Comparing the results of \cite{Mis69b} with our results displayed in FIG.~\ref{fig:trv05-ent}, we indeed find that in the long term, the product $\langle \hat H \rangle\Omega$ stops increasing and tends to remain constant, in fact more and more as the localization of the wavepacket is lost. Our results thus suggest that the results of the semiclassical analysis of \cite{Mis69b} are smoothed out via an infinity of classical trajectories realized at the same time, retaining the constancy of the product $\langle \hat H \rangle\Omega$. In terms of wavepackets, this in turn suggests that the constancy of $\langle \hat H \rangle\Omega$ amounts to a complete loss of localization in the sense that the wavefunction is evenly distributed across the billiard domain.

In the following section, we report the results of our investigations of the quantum cosmological billiards in hyperbolic space. Since the billiard walls are static there, an adiabatic approximation will not be necessary in the first place.

\subsection{Hyperbolic space description}

\begin{figure}[t]
\subfigure[The orbit of FIG.~\ref{fig:cbflat} transformed to the upper half-plane.]{
\includegraphics{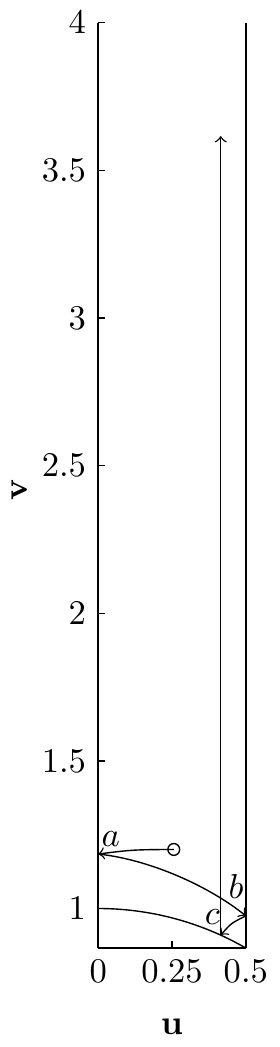}\label{fig:UHP}}
\subfigure[Parametric plot of the trajectory in terms of the expectation values, cf.~FIG.~\ref{fig:UHP}.]{\includegraphics[width=0.5\columnwidth]{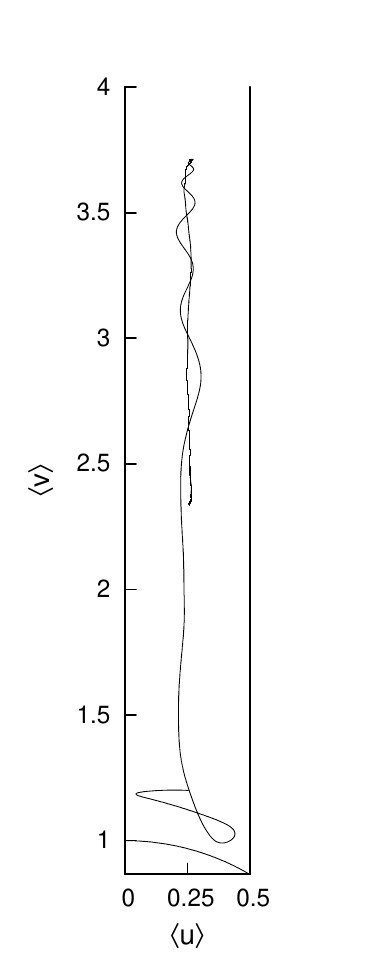}\label{fig:hyp2d-eparam}}
\caption{Classical and wavepacket trajectories in the hyperbolic space description of the quantum cosmological billiards according to initial parameters \eqref{initparams}.}
\vskip1em
\hrule\hrule
\end{figure}

We induce a coordinate transformation to hyperbolic space with $\rho$ as the radial coordinate by setting
\be
\gamma^\mu=\rho^{-1}\beta^\mu \ .
\ee
The cosmological billiard domain of $D=4$ pure gravity is then expressed via
\be
\gamma^1 & = & \frac{(u-\frac{1}{2})^2+v^2+\frac{3}{4}}{\sqrt{3}v} \nn\\
\gamma^2 & = & \frac{(u-\frac{1}{2})^2+v^2-\frac{3}{4}}{\sqrt{3}v} \label{uhptraf}\\
\gamma^3 & = & \frac{\frac{1}{2}-u}{v} \nn
\ee
on the upper half-plane (UHP). The now static billiard walls are accordingly specified by
\be\label{hypwalls}
u\equiv 0\qquad u\equiv \frac{1}{2}\qquad u^2+v^2\equiv 1 \ ,
\ee
as opposed to the moving domain walls in $\beta$-space (cf. FIGS. \ref{fig:cbflat} and \ref{fig:UHP}). The billiard domain on the UHP as specified by \eqref{hypwalls} coincides with the fundamental domain of the group $\PGL_2(\Z)$, which is half the fundamental domain of the modular group $\PSL_2(\Z)$. For numerical investigations of the purely discrete spectrum of odd Maass waveforms for $\PSL_2(\Z)$ and $\PSL_2(\Z[\I])$, we refer to \cite{Ste94,Hej99,The05,AST04}.
The initial parameters \eqref{initparams} of the previous section translate via the relations \eqref{uhptraf} to hyperbolic space and are denoted by $(\rho_0, u_0, v_0)$ in the following.

The initial wavepacket, which is a modulated Gaussian in flat space, is expressed on the hyperbolic plane through
\be\label{hypic}
\Psi(\rho_0,u,v)=A\int_{-\infty}^\infty \dd^2 p\e^{-\frac{c^2}{2}(\vec p - \vec p_0)^2}\e^{\I \rho_0 f(p_+,p_-,u,v)}
\ee
with $f(p_+,p_-,u,v)$ given by
\be
f(p_+,p_-,u,v)= p_+ (\gamma^2(u,v)-\gamma^2(u_0,v_0))  \nn\\
+p_- (\gamma^3(u,v)-\gamma^3(u_0,v_0)) \nn\\
-\omega(p_+,p_-)(\gamma^1(u,v)-\gamma^1(u_0,v_0)) \ .
\ee

The evolution equation for the quantum cosmological billiards is determined by the Wheeler-DeWitt equation with the operator \eqref{WDWophyp} specified to $d=3$, namely
\be\label{WDWhypd3}
\left[\rho^2\ddiffp{}{\rho}+2\rho\diffp{}{\rho}-\LB\right] \Psi(\rho,u,v)=0 \ .
\ee
Here, $\LB$ denotes the Laplace-Beltrami operator
\be
\LB=v^2\left( \ddiffp{}{u}+\ddiffp{}{v}\right)
\ee
on the UHP. As in the previous section, we translate the potential walls into Dirichlet Boundary conditions and study the evolution of the wavepacket numerically, where we implement the differential operators through a finite-differencing scheme using the method of lines, of fourth order in the space-like variables $(u,v)$ as well as in the time-like variable $\rho$ through fourth order Runge-Kutta time integration.

In order to apply the numerical method to the hyperbolic description of the quantum cosmological billiards, the second order Wheeler-DeWitt equation \eqref{WDWhypd3} is split into a system of two first-order PDEs by defining $\Psi_\rho(\rho,x,y)\equiv \rho^2\diffp{}{\rho} \Psi (\rho,x,y)$, leading to the IBVP
\be
\begin{cases}
\diffp{}{\rho}\Psi_\rho(\rho,u,v)=\Delta_{\text{LB}}\Psi(\rho,u,v)\quad  \text{in }\mac{F}\\
\rho^{2}\diffp{}{\rho}\Psi(\rho,u,v) =\Psi_\rho(\rho,u,v)\quad \text{in }\mac{F}\\
\Psi(\rho,u,v)=\Psi_\rho(\rho,u,v)=0 \quad \text{on }\p\mac{F} \\
\Psi(\rho_0,u,v)=f(u,v) ,\ 
\Psi_\rho (\rho_0,u,v)=g(u,v) \ .
\end{cases}
\ee

In addition to \eqref{hypic}, a second initial condition is required. The second one, ${\rho_0}^2(\p_\rho\Psi)(\rho_0,u,v)$, is given by
\be
\I{\rho_0}^2A\int_{-\infty}^\infty\dd^2 p\e^{-\frac{c^2}{2}(\vec p - \vec p_0)^2} f(p_+,p_-,u,v) \times\nonumber\\
\times\left[p_+\gamma^2(u,v)+p_-\gamma^3(u,v)-\omega(p_+,p_-)\gamma^1(u,v)\right] \ .
\ee

For the efficient implementation of the numerical method, we transform the hyperbolic billiard domain (cf.~FIG.~\ref{fig:UHP}) into a rectangle and then compactify it.
Our results have been obtained using a grid size of $1600\times 3920$. They have been subjected to a three-level convergence test including the additional grid sizes $1200\times 2940$ and $800\times 1960$ in order to check the validity of the results.

\begin{figure*}
\subfigure[Initial circular Gaussian wavepacket \eqref{hypic}]{\includegraphics[width=0.66\columnwidth]{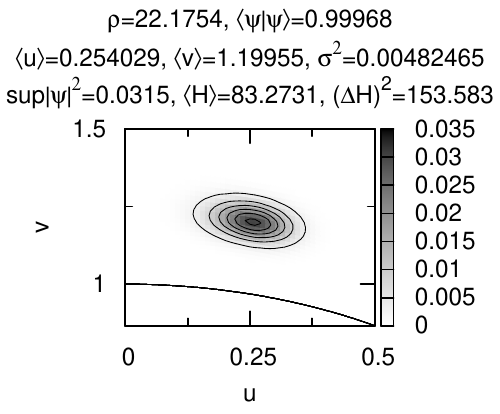}}
\subfigure[Bounce]{\includegraphics[width=0.66\columnwidth]{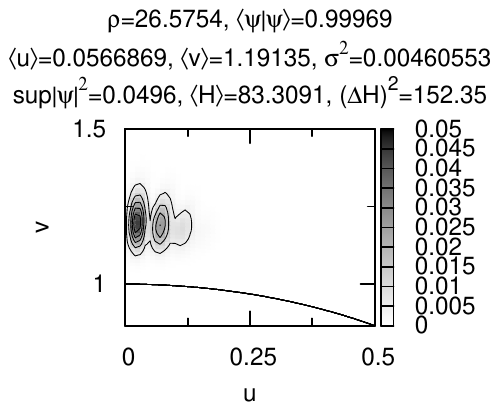}}
\subfigure[Bounce]{\includegraphics[width=0.66\columnwidth]{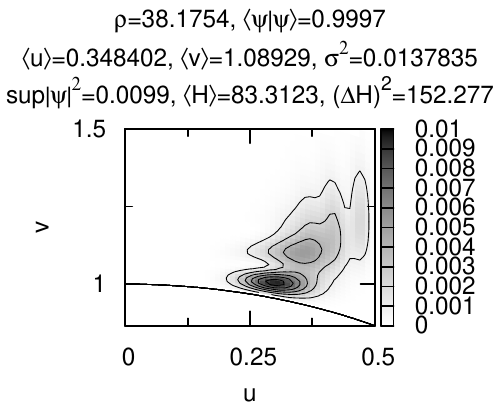}}
\subfigure[Spreading in the direction of $v$]{\includegraphics[width=0.66\columnwidth]{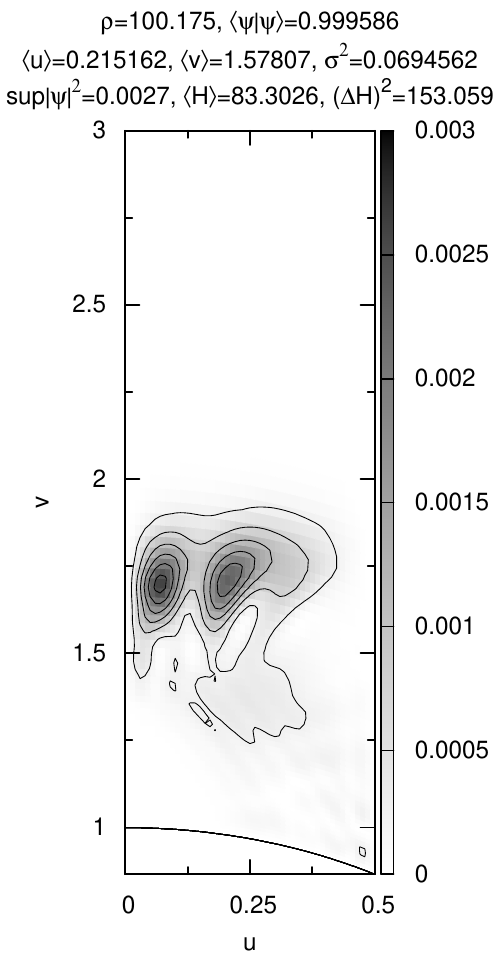}}
\subfigure[Late-time behavior]{\includegraphics[width=0.66\columnwidth]{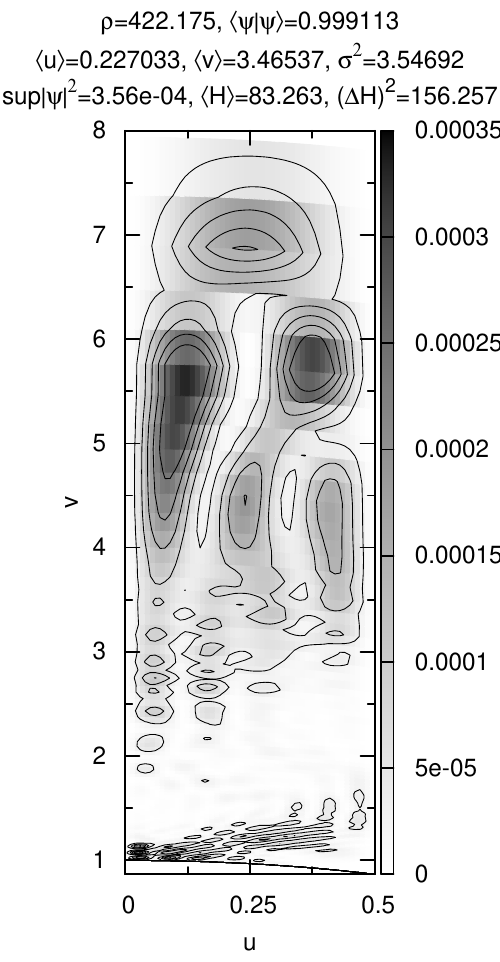}}
\subfigure[Late-time behavior]{\includegraphics[width=0.66\columnwidth]{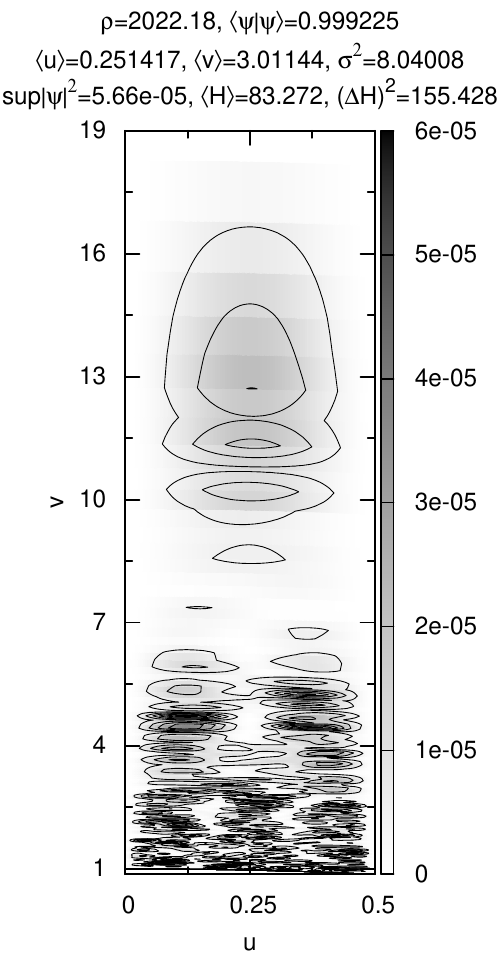}\label{fig:hyp2d-longterm}}
\caption{Snapshots of wavepacket evolution in the quantum cosmological billiards for $D=4$ pure gravity in hyperbolic space.}\label{fig:hyp2d-snap}
\vskip1em
\hrule\hrule
\end{figure*}

\begin{figure*}
\subfigure[The norm of the wavepacket during its evolution according to \eqref{hypscalar}]{\includegraphics[width=\columnwidth]{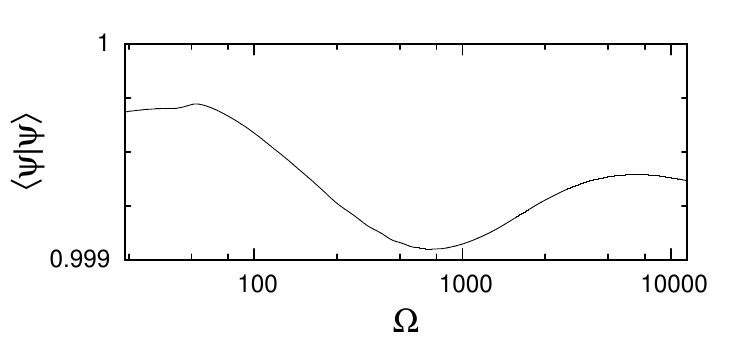}\label{fig:hyp2d-norm}}
\subfigure[Absolute supremum of $|\psi|^2$]{\includegraphics[width=\columnwidth]{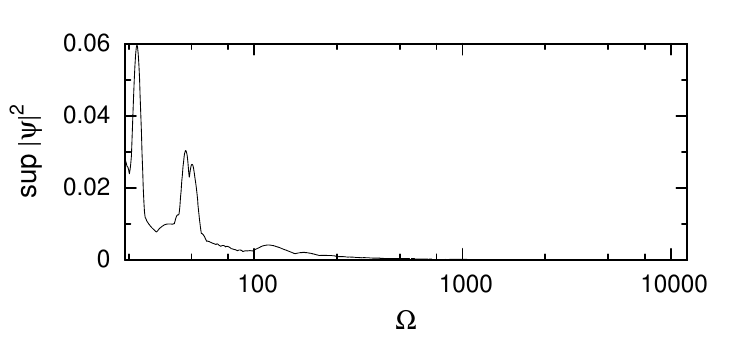}\label{fig:hyp2d-sup}}
\subfigure[Position expectation value on the UHP according to \eqref{posex}]{\includegraphics[width=\columnwidth]{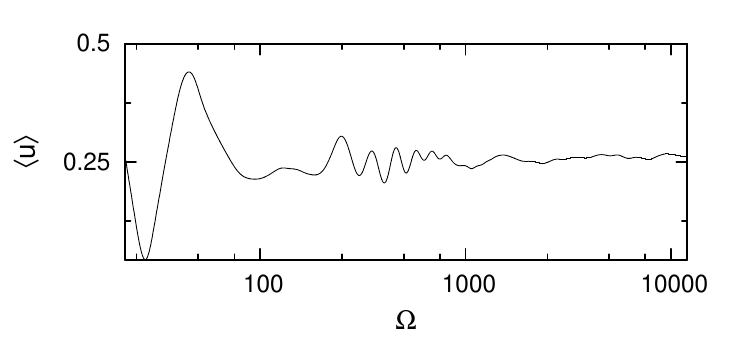}\label{fig:hyp2d-ex}}
\subfigure[Position expectation value on the UHP according to \eqref{posex}]{\includegraphics[width=\columnwidth]{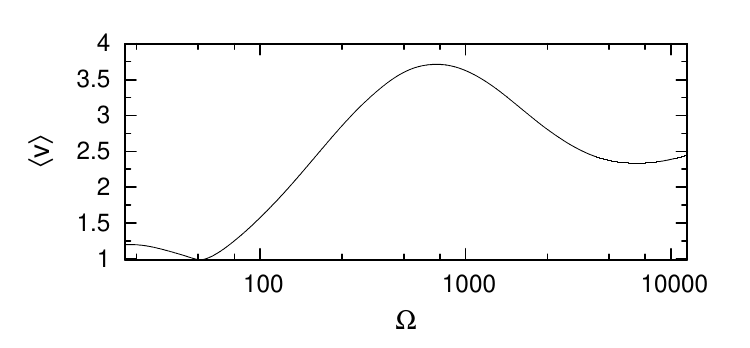}\label{fig:hyp2d-ey}}
\subfigure[Variance of the position expectation value]{\includegraphics[width=\columnwidth]{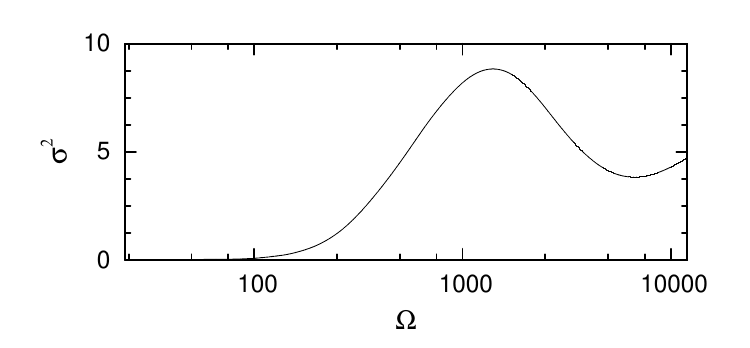}\label{fig:hyp2d-var}}
\subfigure[Energy expectation value of the wavepacket according to \eqref{hypenergy}]{\includegraphics[width=\columnwidth]{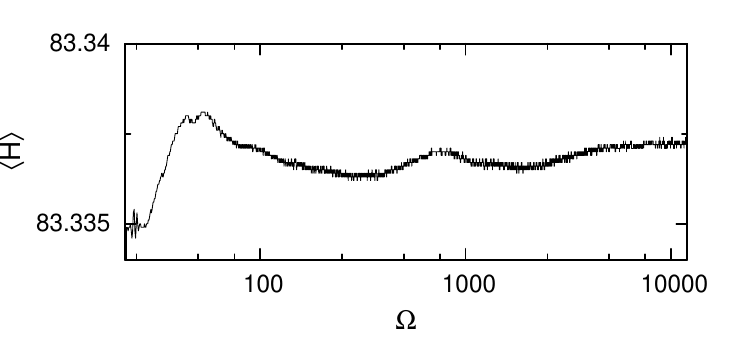}\label{fig:hyp2d-en}}
\subfigure[Variance of the energy expectation value]{\includegraphics[width=\columnwidth]{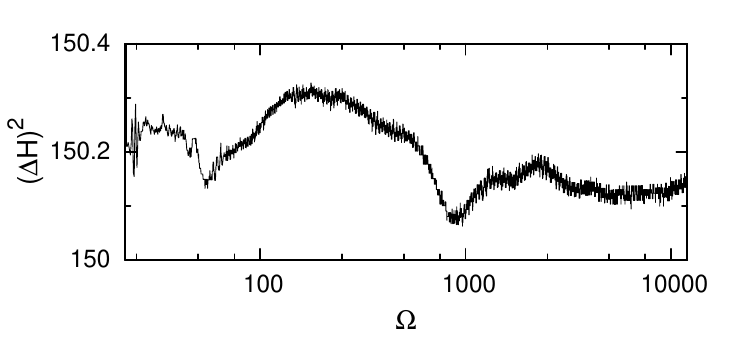}\label{fig:hyp2d-envar}}
\caption{Data for the wavepacket evolution according to initial values \eqref{initparams} in the quantum cosmological billiards for $D=4$ pure gravity in hyperbolic space.}\label{fig:trv05-data}
\vskip1em
\hrule\hrule
\end{figure*}

These results show that, similar to the flat space description, the wavefunction tends to zero pointwise with increasing $\rho$ in the hyperbolic description of the quantum cosmological billiards, cf.~FIG.~\ref{fig:hyp2d-sup}. This suggests the vanishing of the wavefunction at the singularity, matching our predictions of \cite{KKN09}, see also equation \eqref{WDWsolrad}. The position expectation values according to \eqref{posex} of the wavepacket's evolution in the billiard domain on the hyperbolic UHP  are displayed in FIGS.~\ref{fig:hyp2d-ex} and \ref{fig:hyp2d-ey}. The parametric plot of the position expectation values displayed in FIG.~\ref{fig:hyp2d-eparam} shows that the wavepacket follows at first the classical trajectory shown in FIG.~\ref{fig:UHP} and then starts to deviate from it as it is spreading out. After the last reflection at point $c$, the classical path depicted in FIG.~\ref{fig:UHP} is following a straight ``vertical'' line towards the cusp. Since on the UHP every geodesic with respect to the Poincar\'e metric is either a half-circle or a vertical straight line, all the neighboring rays representing the wavepacket are given by half-circles. However, all these rays are subject to reflections from the billiard walls and will only reach a finite height with respect to the $v$ coordinate. This is reflected in the wavepacket's trajectory displayed in FIG.~\ref{fig:hyp2d-eparam} which has a turning point at a certain value of $v$, contrary to the classical path. Contour plots of the evolution of the wavepacket are shown in FIG.~\ref{fig:hyp2d-snap} at different instants of $\rho$, chosen such as to display similar situations as in FIG.~\ref{fig:trv05-snap}.

The energy expectation value and its variance described in the hyperbolic description by \eqref{hypenergy} converge with increasing numerical resolution to a constant value throughout the wavepacket's evolution, cf.~FIGS.~\ref{fig:hyp2d-en} and \ref{fig:hyp2d-envar}. We pick up where we have left at the end of the preceding section and compare our results with the semiclassical considerations of \cite{Mis69b}. In the hyperbolic description of the cosmological billiards, the billiard walls are static and thus we do not have to rely on an adiabatic approximation, contrary to the description in flat space of the previous section. Since the energy expectation value and its variance remain constant, we directly see that a highly excited semiclassical state stays highly excited on its way ``into'' the singularity as $\rho\rightarrow\infty$.
On the other hand, we observe that the localization of the semiclassical wavepacket is lost due to the spreading across the billiard domain, and particularly that the wavefunction tends to zero as the singularity is approached. In \cite{Mis69b} it was deduced that the quantum state remains classical all the way into the singularity. In the current framework however, we propose to rather take the vanishing of the wavefunction due to its spreading (and thereby its loss of localization) than the excitation in terms of energy levels as an indicator for a quantum resolution of the singularity.

We conclude this section by remarking that the wavepacket evolution in hyperbolic space can be mapped back to flat space in order to double-check the correctness of the computation. The set of hyperboloids which make up the discretization of the forward lightcone has to be sliced into equal-$\Omega$-time hypersurfaces, such that for every such ``horizontal'' slice, there is a one-dimensional contribution from each intersecting hyperboloid. Using interpolation, one can then confirm that the evolution under \eqref{WDWhypd3} with the initial conditions \eqref{hypic} indeed matches the evolution under \eqref{KGqcbflat} with the initial conditions \eqref{initparams} that has been reported in the previous section.

\section{Summary and discussion}
\vspace{-0.27em}
Pure gravity and supergravity in four to eleven spacetime dimensions admit a description in terms of cosmological billiards in the BKL limit towards a spacelike singularity. These singularities of the classical theory are expected to be resolved through quantum effects. The classical billiard motion of the BKL limit is classified as strongly chaotic for theories such as four-dimensional pure gravity and eleven-dimensional supergravity, and exact solutions for the quantization of these classically chaotic billiard systems are not available. With this article, we supplement and extend our previous results \cite{KKN09} by reporting investigations regarding the evolution of relativistic wavepackets towards the singularity as the quantum cosmological billiards associated with $D=4$ pure gravity. In agreement with our previous results \cite{KKN09}, we found that the wavefunction of the universe decreases pointwise towards zero as the singularity is approached, while its norm is preserved.

The initially localized wavepacket evolves according to the Wheeler-DeWitt equation, which reduces to a massless Klein-Gordon equation for the finite-dimensional cosmological billiards. The subtleties of relativistic wavepackets in domains with moving walls have been elucidated first in the one-dimensional case in section \ref{sec:1d}, and an exact set of solutions has been derived, for the massless as well as for the massive relativistic Klein-Gordon quantum particle inside the infinite square well with moving walls. Since analytic solutions for the quantum cosmological billiards associated with $D=4$ pure gravity are however not known and out of reach, we investigated the relativistic wavepacket evolution using numerical methods in section \ref{sec:qcb}. Two different realizations have been employed. The first one is in flat space with moving billiard walls, while the second one is in hyperbolic space with static walls. The solutions in the two realizations for corresponding choices of initial parameters can be mapped into each other, and in both cases tend towards zero with preserved norm as the singularity is approached.

In the flat space description of the cosmological billiards, the position (i.e.~anisotropy) expectation values follow at first the corresponding classical trajectory and then start to deviate as the wavepacket spreads across the billiard domain with increasing variance due to transversal fanning and reflections off the moving walls. Upon such reflection off a moving wall, the wavepacket loses energy into the wall and gets redshifted.  As in the flat space description, the position expectation values of the wavepacket in the hyperbolic space description of the quantum billiards follow at first the classical path before deviating due to the spreading. A classical particle can travel on a geodesic all the way into the cusp of the fundamental domain at infinity, whereas the wavepacket trajectory in terms of its position expectation values necessarily has a turning point. The corresponding sections of this article have each been supplemented with a set of snapshots of the evolution of the wavepacket. These results can be contrasted to the quantum billiards of relativistic wavepackets in a static triangle, reported in the appendix.

We furthermore observe that the product of energy expectation value and time tends to a constant value in the flat space description, in agreement with the results of the semiclassical analysis of \cite{Mis69b} which imply that a highly excited semiclassical state remains highly excited and thus classical all the way into the singularity. However, the preserved excitation becomes more obvious in the hyperbolic space description of the cosmological billiards, where the billiard domain of $D=4$ pure gravity happens to coincide with the fundamental domain of the group $\PGL_2(\Z)$. There, the billiard walls are static, making an adiabatic approximation redundant, and we found that the energy expectation value and its variance are preserved. Nevertheless, regardless of how highly excited in terms of eigenstates, the loss of localization of the initial wavepacket as it spreads across the billiard domain, and particularly the vanishing of the wavefunction suggest the interpretation that the singularity is resolved via an effective disappearance of spacetime near the singularity, making it effectively unreachable, possibly in line with a ``deemergence'' in terms of a Kac-Moody algebra coset model description \cite{DN08}. In this light, we thus propose that the vanishing of the wavefunction and its spreading across the whole billiard domain may be considered as more appropriate indicators for a quantum resolution of the classical singularity than the excitation of the quantum state in terms of energy levels.

It should be remarked at this point that the quantum mechanical avoidance of classical cosmological singularities has also been discussed in terms of approaches that are different from the one employed here. In some of the models recently investigated in \cite{BLKSVM09}, vanishing of the wave function near the classical singularity is found, too. In general, the resolution of the singularity that is suggested by the vanishing of the wavefunction is to be contrasted to other mechanisms such as the Hartle-Hawking no-boundary proposal \cite{HH83} or the cosmic bounces of loop quantum cosmology (see \cite{AS11} for a recent review), since the latter two require continuation of the wave packet into the singularity and beyond. It should also be noted that there exist approaches that circumvent the assumptions of the singularity theorems of Hawking and Penrose in the first place, for example through a violation of the null energy condition from a ghost condensate \cite{AHCLM04}. Such a mechanism can produce a smooth, non-singular bounce where an ekpyrotic contracting phase is joined with the standard expanding phase \cite{LMTS07,BKO07,CS07,Leh11}, thereby providing new solutions for cosmological issues such as the flatness problem.

The setting considered in this article only provides a very first step. Albeit very useful, the cosmological billiards scenario with its infinite potential walls represents only crudely the spatial inhomogeneities and possible matter degrees of freedom of the theory. Since the shape of the billiard table points to an underlying structure in terms of Lorentzian Kac-Moody algebras, the quantum state of the universe would have to be generalized according to the ``small tension'' expansion of \cite{DHN02} as the ensuing first step away from the BKL limit of the classical theory. We conclude this article by emphasizing that it is conjectured that in the deep quantum regime, spacetime based quantum field theory is replaced by a coset model built from the respective Kac-Moody algebra that defines the quantum cosmological billiards \cite{DHN02,DN08}.


\acknowledgments{
I would like to thank Hermann Nicolai, Christian Reiss\-wig, Holger Then, William Schiesser, Ian Hinder, Margaret Hawton, Jakob Palmkvist, Axel Kleinschmidt, Stefan Pfenninger, Burkhard Schwab, Philipp Fleig, Parikshit Dutta and Florian Loebbert for helpful discussions.
}

\appendix

\section{The triangular quantum billiards with static domain walls}\label{app}

\begin{figure*}[!t]
\subfigure[At $\Omega=9884$, the initially localized wavepacket has spread over the static triangular billiard domain.]{\includegraphics[width=\columnwidth]{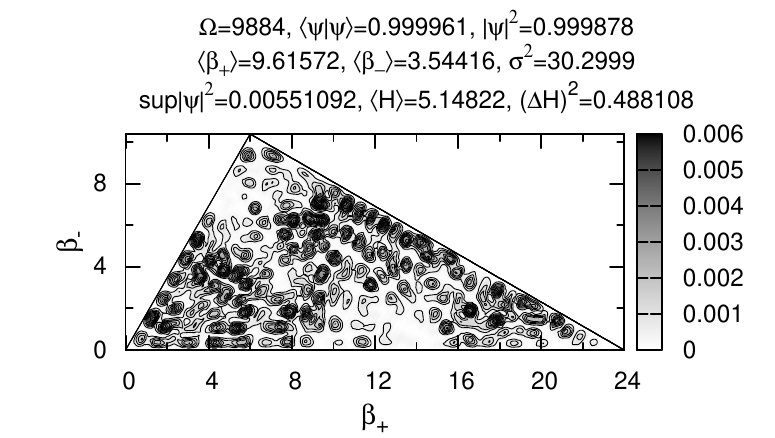}}
\subfigure[Supremum norm]{\includegraphics[width=\columnwidth]{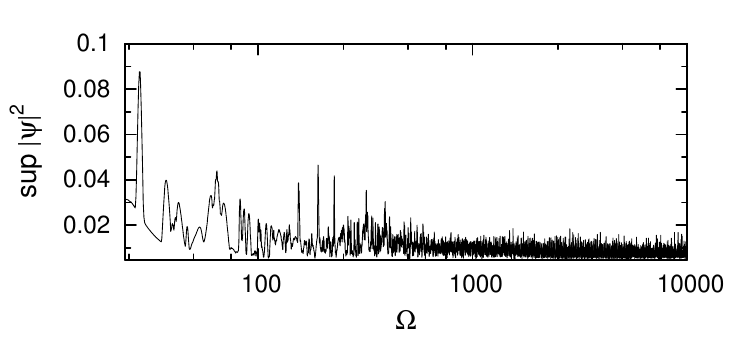}}
\subfigure[Variance]{\includegraphics[width=\columnwidth]{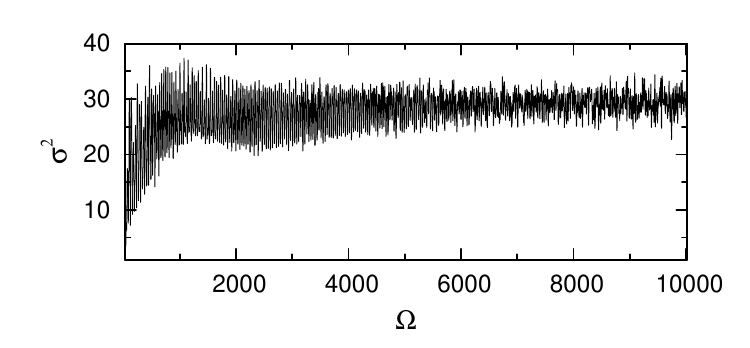}}
\subfigure[Parametric plot of expectation values trajectory]{\includegraphics[width=\columnwidth]{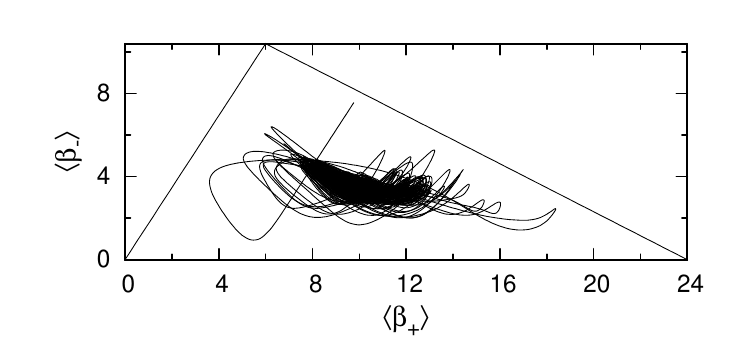}}
\subfigure[Expectation value for $\bp$]{\includegraphics[width=\columnwidth]{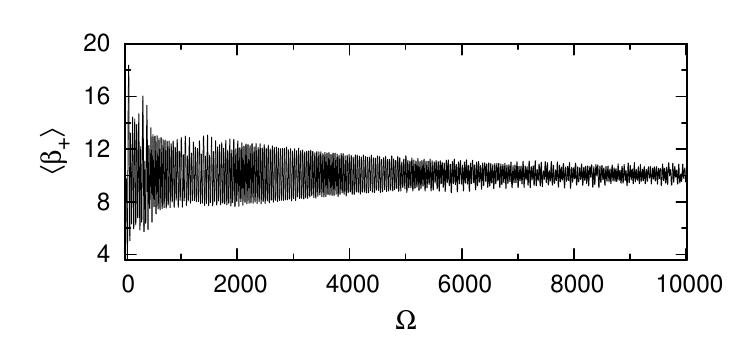}}
\subfigure[Expectation value for $\bm$]{\includegraphics[width=\columnwidth]{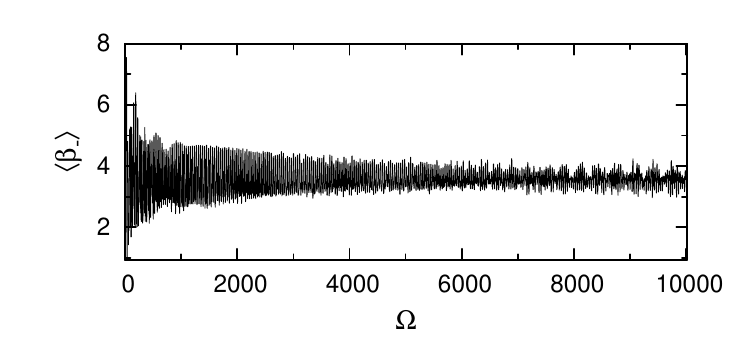}}
\subfigure[Energy expectation value]{\includegraphics[width=\columnwidth]{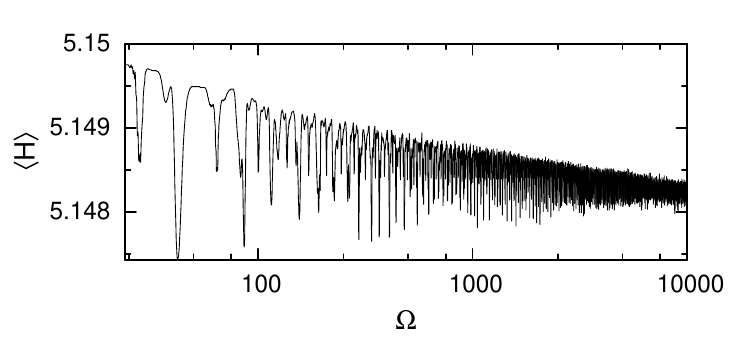}}
\subfigure[Variance of energy expectation value]{\includegraphics[width=\columnwidth]{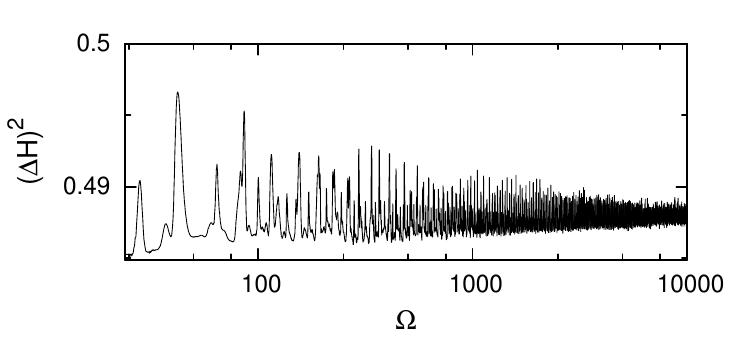}}
\caption{The quantum billiards in the nonintegrable static triangle that corresponds to the initial cosmological billiard domain from the previous section with each billiard wall kept fixed. The norm $\langle \Psi|\Psi\rangle\approx|\Psi|^2\approx 1$ during the whole evolution, oscillating only within bounds of $5\cdot 10^{-6}$. The grid size was $800\times 800$ data points.}\label{fig:trv0-s3}
\vskip1em
\hrule\hrule
\end{figure*}

In this appendix, we study the triangular quantum billiards with the same parameters as in the quantum cosmological billiards before, but this time with every single one of the billiard walls remaining static from the outset. In this case, the corresponding classical billiards is completely integrable. Note that for triangular billiards with static walls, only three cases are classically completely integrable, namely the equilateral triangle, the half-equilateral triangle, or the half-square. These are the cases that feature a second constant of the motion, and the invariant surface of the phase flow is given by a two-dimensional torus in the four-dimensional phase space. Despite the non-separability of the triangular billiards, analytic solutions are indeed known for the quantum billiards that correspond to these classically integrable cases \cite{Lam52,Pin80,KMV82,Li84}. For a right triangle with rational angles other than the two cases mentioned above, the classical billiards is only pseudointegrable, and although its trajectories are still confined to a two-dimensional surface in phase space, this surface is not a torus but has higher genus \cite{RB81}, and analytic solutions for the corresponding quantum billiards are not known. We also note that for a right triangle with non-rational acute angles, there exists no other constant of motion besides the Hamiltonian and the classical motion is ergodic and fills the whole phase space.

We set the parameters as in \eqref{initparams} and investigate the wave packet propagation numerically. As in the calculations reported in the previous sections, a three-level convergence test was applied on the numerical computations. A snapshot of the long-term behavior is plotted in FIG.~\ref{fig:trv0-s3}. The Klein-Gordon norm according to \eqref{KGscalar} remains constant within bounds of $5\cdot 10^{-6}$, and as expected, the variance now has an upper and the supremum norm a lower bound. The position expectation values according to \eqref{posex} are oscillating around the center of mass of the triangle. The energy expectation value according to \eqref{energy1d} converges with increasing resolution to a constant value, and so does its variance.

\bibliography{koehn-rwiccqcb}

\end{document}